\documentstyle[12pt]{article}

\newcommand{\be}{\begin{equation}}
\newcommand{\ee}{\end{equation}}
\newcommand{\bea}{\begin{eqnarray}}
\newcommand{\eea}{\end{eqnarray}}

\begin{document}
\begin{titlepage}

\begin{flushright}
{\today}
\end{flushright}
\vspace{1in}

\begin{center}
\Large
{\bf Proof of Dispersion Relations for the 
Amplitude  in Theories with a Compactified Space Dimension    }
\end{center}

\vspace{.2in}

\normalsize

\begin{center}
{ Jnanadeva Maharana \footnote{Adjunct Professor, NISER, Bhubaneswar}  \\
E-mail maharana$@$iopb.res.in} 
\end{center}

\normalsize

\begin{center}
 {\em Institute of Physics \\
Bhubaneswar - 751005, India  \\
   and \\
   Max-Planck Institute for Gravitational Physics, Albert Einstein Institute, Golm, Germany  }

\end{center}

\vspace{.2in}

\baselineskip=24pt

\begin{abstract}

The analyticity properties of the scattering amplitude  in the nonforward direction are investigated for a field theory in the manifold
$R^{3,1}\otimes S^1$.  A scalar field theory of mass $m_0$ is considered in $D=5$ Minkowski space to start with. Subsequently, 
one spatial dimension is compactified to a circle; the $S^1$ compactification.
 The mass spectrum of the resulting theory is: (a) a massive scalar of mass, $m_0$, same as the original  five dimensional theory  and (b) 
 a tower of  massive  Kaluza-Klein states. We derive nonforward dispersion relations for scattering of the excited Kaluza-Klein states
in the Lehmann-Symanzik-Zimmermann formulation of the theory. 
 In order to accomplish this object, first we generalize  the Jost-Lehmann-Dyson theorem for a relativistic field theory with a compact spatial
 dimension. Next, we show the existence of the Lehmann-Martin ellipse inside which the partial wave expansion converges. It is proved that 
 the  scattering amplitude satisfies fixed-t dispersion relations when $|t|$  lies within the Lehmann-Martin ellipse.

\end{abstract}

\vspace{.5in}

\end{titlepage}


\noindent {\bf 1. Introduction}

\bigskip
\noindent
This article is a continuation of our investigation of the  analyticity properties of scattering amplitude in scalar field theory  defined in a manifold
$R^{3,1}\otimes S^1$. First we consider a neutral, massive scalar field  theory of mass $m_0$  in a flat five dimensional Minkowski space. Subsequently, one spatial
coordinate is compactified on a circle of radiius $R$.  The spectrum of the resulting theory consists of a neutral scalar of mass $m_0$ 
(same as the  mass of the 
original uncompactified theory) and a tower of massive Kaluza-Klein (KK) states carrying the KK charges. We adopt the Lehmann-Symanzik-Zimmermann (LSZ)
\cite{lsz} formalism to construct the amplitude and to study the analyticity property of the scattering amplitude. We had proved the forward
dispersion relation for scattering of KK states in an earlier paper \cite{jm18} ( henceforth referred to as I).  The present investigation brings our
programme to a completion.\\ 
The analyticity properties of scattering amplitude plays a very important role in our understanding
of collisions of relativistic particle in the frame works of general field theories without appealing to any specific model. 
The scattering amplitude, $F(s,t)$,  is an
analytic function of  the center of mass energy squared, $s$,   for fixed momentum transferred squared, $t$. 
 The fixed-t  dispersion relations in $s$ have been proved
when $|t|$  lies within the Lehmann  ellipse in the axiomatic approach in the case of $D=4$ field theory, mostly 
for a single neutral massive field.. These results are derived from the general field theories (axiomatic field theories) in the 
the axiomatic    approach of
Lehmann-Symanzik-Zimmenmann (LSZ) \cite{lsz} and  in the more general frameworks of axiomatic formulation of field theories
\cite{book1,book2,book3, fr1,lehm1,sommer,eden,wight,jost,streat,kl,ss,bogo}.
We recall that  some of the fundamental principles of such formulations are  locality, mircocausality, Lorentz invariance to mention a few.
 There are very strong reasons to believe  that if the dispersion relations are violated then  the validity of some of axioms of
 these generalized relativistic field theories might be in question.  The subsequent progress in this field has led to several rigorous
  theorems which impose constraints on experimentally observable parameters, generally stated as bounds. These bounds
  have been put to tests in high energy collision experiments and there is so far no evidence of the violation of these bounds. Notable
  among them is the  Froissart-Martin bound \cite{fr,andre, andre1} that restricts the growth of total cross sections at asymptotic energies;
  \bea
  \label{fmb}
  \sigma_t\le {{4\pi}\over{t_0}}(log s)^2
  \eea
   where $t_0$ is determined from the first principles for a given scattering process. The experimental data
  respect this upper bound for numerous  scattering processes over a wide range of energy.
  In the event of any experimental violations of the bound, we shall be compelled to reexamine some of the axioms of the general theories. 
\\
The scattering amplitude in nonrelativistic potential scattering
exhibit certain analyticity properties in energy $k$ for a large class of potentials as is known for a very long time  
 \cite{gw,khuri1,wong}. We recall that that the analyticity of scattering amplitude in QFT enjoys a very intimate relationship
 with the principle of microcausality. In contrast, however, in the context of potential scattering, there is no such
 deep reason which leads to analyticity of the corresponding amplitude.
 We recall that the  nonrelativistic theory is  
invariant only under Galilean transformations whereas QFT's are required to be Lorentz invariant.   Khuri \cite{khuri2} encountered a situation, in a 
nonrelativistic potential model, where the amplitude
does not satisfy analyticity in momentum $k$. The consequences of such a violation of analyticity  would not the
so serious for scatterings in a nonrelativistic potential model.
 Whereas, if the amplitude constructed in the frame works of general field theories based on LSZ or Wightman axioms,
does not exhibit analyticity then it will raise serious concerns.
\\
The roles played by higher spacetime dimensional field theories ($D>4$) has become increasingly important.
One of the primary reasons is that our quest to construct unified fundamental theories have led physicists
to explore consistent theories in higher spacetime dimensions so that the physical phenomena understood
in four spacetime dimensions are through effective theories. 
It is worth while to recall, in this context,  supersymmetric theories, supergravity theories and the string theories which have been
investigated intensively over past several decades, are consistently defined in higher spacetime dimensions. 
 In order to understand the physics in four dimensions, we adopt the ideas
of Kaluza-Klein compactifications in the modern perspective. Thus it is invoked that some of the extra spatial dimensions are
compactified in order to facilitate construction of four dimensional theories enabling us to comprehend physical phenomena
observed in the present accessible energies.
There are a large class effective four dimensional theories arising from various compactification schemes. Moreover, there are proposals, the
so called large radius compactification schemes where  the signatures of the extra spatial dimensions might be
observed in current high energy colliders \cite{anto,luest}.  
As a consequence, there has been a lot of
phenomenological studies to investigate and build models for possible experimental
observations of the decompactified dimensions at the present high energy accelerators such as LHC.
 These two review articles \cite{anto,luest} explain the large radius compactification
proposal in detail; they contain an extensive list of articles on the topic. 
Indeed,  the scale of the extra compact dimensions is extracted from the  LHC experiments and it puts the 
compactification scale to be more than  2 TeV \cite{tev1,tev2}.  
The signatures of models of large radius
compactification and the number of extra compactified dimensions envisaged in a model,  go into getting the
experimental limits. In some cases, even the limit could be higher than 2 TeV and we refer the readers to the two
papers cited here. \\
The large radius compactification ideas motivated Khuri \cite{khuri2}, in order to investigate the analyticity properties of
scattering amplitude in a nonrelativistic potential model.   He identified a model where the potential is spherically
symmetric as a function of noncompact coordinates and is of short range, on the other hand one extra spatial coordinate
is compctified on $S^1$.  Khuri \cite{khuri2} 
 discovered that, under certain circumstances,  the amplitude does not always satisfy the analyticity  properties. He also recalled that  the analyticity
properties  of amplitudes were investigated earlier  \cite{khuri1,wong} with noncompact spatial coordinates  (for $d=3$ case; 
 when  there was no $S^1$ compactification)
the amplitude satisfied the dispersion relations.
  Khuri \cite{khuri2} provided counter examples for a model with the  $S^1$  compactification  to demonstrate 
  how the analyticity of the forward scattering amplitude
  beaks down in the presence of $S^1$ compactification. This result is based on perturbation theoretic approach to nonrelativistic potential
  scattering. We shall very briefly summarize Khuri's result in the next section. \\
  It was  shown in (I) that the forward scattering amplitude in a relativistic quantum field theory (QFT), with a compact spatial
  coordinate,  satisfies forward
  dispersion relation unlike what Khuri had concluded in his potential model \cite{khuri2}. We had considered a five dimensional massive, neutral scalar field theory in five dimensional Minkowski (flat)
  space to start with. Subsequently, one spatial coordinate was compactified on $S^1$. The LSZ formalism was adopted to
  derive the scattering amplitude. As mentioned earlier, if dispersion relations were violated in such a theory then foundations
  of general relativistic quantum field theories would be questioned. However, the proof of dispersion relations in the forward
  direction does not provide a complete study of analyticity properties of the theory. It is necessary to prove the nonforward dispersion
  relations for a general  relativistic QFT.  We had discussed the requisite steps necessary in order to accomplish this goal in I. 
  The purpose this article is to bring to completion  the investigation of the analyticity of the four point amplitude.
  \\
  We recall our previous work \cite{jmjmp1} on study of analyticity in higher dimensional theories as those results will be quite
  useful for the continuation to the present investigation.     
   We proceeded as follows to study high energy behaviors and analyticity of higher dimensional theories \footnote{  The usefulness
   of our work has been noted by Du Lecroix, Erbin and Sen \cite{sen} and collaborators in their string field theory approach to prove crossing symmetry for scattering
   of stringy states}.
     It was shown, in the LSZ formalism,  that
the scattering amplitude has desire attributes in the following sense: (i)  We proved the generalization of the
Jost-Lehmann-Dyson theorem for the retarded function \cite{jl,dyson} for the $D>4$  case \cite{jmplb}. (ii) Subsequently, we
  showed the existence of the
Lehmann-Martin ellipse for such a theory. (iii) Thus a dispersion relation can be written in $s$ for fixed $t$
when the momentum transfer squared lies inside Lehmann-Martin ellipse \cite{leh2,martin1}. (iv) The analog of Martin's theorem can be
derived in the sense that the scattering ampliude is analytic the product domain $D_s\otimes D_t$ where $D_s$ is the 
cut $s$-plane and $D_t$ is a domain in the $t$-plane such that the scattering amplitude is analytic inside a disk, $|t|<{\tilde R}$, ${\tilde R}$ is radius of the
disk and it is  independent of
$s$.  Thus the partial wave expansion converges inside this bigger domain. (v) We also derived the analog of Jin-Martin \cite{jml} upper bound on the scattering amplitude which states
that the fixed $t$ dispersion relation in $s$ does not require more than two subtractions. (vi)  Consequently, a generalized
Froissart-Martin bound was be proved.\\
In order to accomplish our goal for a $D=4$ theory which arises from $S^1$  compactification of a $D=5$ theory i.e. to prove nonforward 
dispersion relations, we have to establish the results (i) to (iv) for this theory. It is important to point out, at this  juncture, that (to be elaborated in sequel)
the spectrum of the theory consists of a massive particle of the original five dimensional theory and a tower of Kaluza-Klein states.
Thus the requisite results (i)-(iv) are to obtained in this context in contrast to the results of the D-dimensional theory with a single massive neutral scalar
field. \\
The paper is organized as follows.  In the next section (Section 2)  we recapitulate the main results of Khuri's work  \cite{khuri2}
without details. The interested reader might consult the original paper of Khuri or section 2 of paper I.
We also provide a brief account of I without details. Thus the notations, conventions and main results will be presented here.
The third section is devoted to investigation of the analyticity of the scattering amplitude. Our first step is to obtain the Jost-Lehmann-Dyson
representation. Consequently, we would obtain the domain free from singularity in $t$-plane. Next, we shall outline the derivation of the
Lehmann ellipses in the present context. The derivation needs to account for the fact that, unlike the case of  the usual derivation for  single scalar
theory, there are the KK towers whose presence has to be taken into considerations. Subsequently, we are in a position to write the nonforward
dispersion relations.
In Section 4 we explore the consequences of unitarity. In order to complete our studies, we have to invoke the unitarity of 
S-matrix as will be clarified later. One interesting result is that in derivation of the spectral representation of vacuum expectations values of the $R$-product
of currents the tower of KK states are to be included when we sum over the complete set of 
intermediate physical states. The unitarity conditions impose constraints on contributions of the intermediate physical states.
The second accomplishment  is to obtain a result analogous Martin's theorem where unitarity and consequently, the positivity properties
of the partial wave amplitudes are important. We also derive a version of the  Froissart-Martin bound as a corollary
for a field theory with $S^1$ compactification.
Another  important corollary that follows from this study is that the nonforward amplitude, when $|t|$ is within the Lehmann-Martin ellipse, requires
at most two subtractions in the fixed-$t$ dispersion relations; thus deriving a bound analogous to the Jin-Martin bound.  We summarize 
the contents of this work and discuss
our results in Section 5.
 
\newpage

\noindent {\bf 2. Analyticity Property of Scattering Amplitude and Compact Spatial Dimension. }\\
 
In this section, we  shall shall briefly present some of the results which motivated the present investigation. We enlist important axioms and
the relevant kinematical variables.
 First we summarize essential results of  Khuri's work \cite{khuri2}. The interested reader on this topic
may go through his paper for details.

\bigskip
\noindent{\bf 2.1.  Scattering in nonrelativistic Quantum Mechanics with a Compact Dimention}

\bigskip

\noindent Khuri \cite{khuri2} studied analyticity property of scattering amplitude in a nonrelativistic potential  model with
a compact spatial dimension. The theory is defined as follows: the potential is $V(r,\Phi)$, where $r$ is the radial coordinate,
$|{\bf r}|=r$, of the three dimensional space and $\Phi$ is compact coordinate; $\Phi+2\pi R=\Phi$. 
 The radius of compactification, $R$, is  taken to be very small, $R<<1$, compared to the scale available in the potential
 theory (there is no Planck scale here). The perturbative
Greens function technique is adopted. The scattering amplitude depends on three variables: the momentum $\bf k$, scattering angle and an
integer associated with the periodicity of $\Phi$. The free Greens function satisfies the free Schr\"odinger equation: 
\bea
\label{kh5}
\bigg[{ \nabla}^2+{{1\over R^2}}{{\partial}^2\over{\partial\bf\Phi}^2} +K^2\bigg] G_0({\bf K};{\bf x},\Phi:{\bf x'},\Phi ')=\delta^3({\bf x}-{\bf x'})\delta(\Phi -\Phi')
\eea
The plane wave solution to the Schr\"odinger equation is $\Psi_0({\bf x}, \Phi)={1\over{(2\pi)^2}}e^{i{\bf k}.{\bf x}}e^{i n\Phi}  $, $n\in {\bf Z}$
and $K^2=k^2+(n^2/R^2)$ . The closed form expression for the free Greens function has been derived  in \cite{khuri2}. A notable feature
is that for  $(n^2/R^2)> K^2$ the Greens function is exponentially dampted as $e^{-{\sqrt{n^2/{R^2}-K^2}}}$.\\
The expression for the scattering amplitude is extracted from the large $|{\bf x}|$ limit when one looks at the asymptotic behavior of 
the wave function, 
\bea
\label{kh9}
\Psi_{{\bf k},n}\rightarrow e^{{\bf k}.{\bf x}}e^{in\Phi}+ \sum_{m=-[KR]}^{+[KR]}T({\bf k'},m:{\bf k},n){{e^{ik'_{mn}|{\bf x}|}\over{|{\bf x}|}}}e^{im\Phi}
\eea
where $[KR]$ is the largest integer less than $KR$ and 
\bea
\label{kh10}
k'_{mn}={\sqrt{k^2+{{n^2}\over{R^2}}-{{m^2}\over{R^2}}}}
\eea
Khuri \cite{khuri2}  identifies a conservation rule:
 $K^2=k^2+(n^2/{R^2})=k'^2+(m^2/{R^2})$. Moreover, it is argued   that the scattered wave has only $(2[KR]+1)$ components and those states with
$(m^2/(R^2)>k^2+(n^2/{R^2})$ are exponentially damped for large $|{\bf x}|$ and consequently,  these do not appear in the scattered wave. 
Now the scattering amplitude is extracted by Khuri using the standard prescriptions. It  takes the following form  
\bea
\label{kh11}
T({\bf k'},n';{\bf k},n)=-{{1\over{8\pi^2}}}\int d^3{\bf x'}\int_0^{2\pi} d\Phi'e^{-{\bf k'}.{\bf x'}}e^{-in'\Phi'}V({\bf x'},\Phi')\Psi_{{\bf k},n}({\bf x'},\Phi')
\eea
Note that the condition,  $k'^2+n'^2/{R^2}=k^2+n^2/{R^2}$ is to be satisfied. Thus the scattering amplitude describes the process where incoming wave
$|{\bf k},n>$ is scattered to final state $|{\bf k'},n'>$ with the above constraint.\\
Khuri proceeds further to extract the scattering amplitude starting from the full Greens function. It satisfied the Schr\"odinger equation in the
presence of the potential. The equation assumes the following form
\bea
\label{kh12}
T({\bf k'},n';{\bf k},n)-T_B=&&-{{1\over{8\pi^2}}}\int ....\int d^3{\bf x}d^3{\bf x'} d\Phi d\Phi'e^{-i({\bf k'}.{\bf x'}+n'+\Phi'}
V({\bf x'},\Phi')\nonumber\\&&G({\bf K};{\bf x'},{\bf x};\Phi',\Phi) V({\bf x},\Phi)e^{i({\bf k}.{\bf x}+n\Phi)}
\eea
Here $T_B$ is the Born term given by
\bea
\label{kh12a}
T_B=-{1\over{8\pi^2}}\int d^3x\int_0^{2\pi}e^{i({\bf k}-{\bf k}).{\bf x}}V(x,\Phi)e^{i(n-n')\Phi}d\Phi
\eea
The perturbative Greens function technique is utilized  to extract  the scattering amplitude order by order. The crucial observation of Khuri \cite{khuri2}
is that when he considers the forward amplitude for the case of $n=1$,  to second order, the amplitude does not satisfy analyticity property in $k$, whereas
for $n=0$ he does not encounter any such problem. 
He had considered a general class of potentials of the type
\bea
\label{kh14}
V(r,\Phi)=u_0(r)+2\sum_{m=1}^N u_m(r)cos({m\Phi})
\eea
  where $u_m(r)=\lambda_m{{e^{-\mu r}} \over{ r}}$ and the potential is short range in nature.
   Khuri drew attention to an important fact that in absence of any compactified
  coordinates, when analyticity of scattering amplitude was investigated in a theory in the 3-dimensional space with same type of potential as above, 
  the amplitude did respect analyticity \cite{khuri1,wong}.\\
  {\it Remarks}: (i) Khuri \cite{khuri2} noted that, in the context of large radius compactification scenario, if the amplitude exhibits such
  a nonanalytic behavior in $k$, there will be serious implications for the physics at LHC energies.\\
  (ii) Moreover, it is to be noted that in the frame works of nonrelativistic quantum mechanics, the analyticity of scattering amplitude
  is not so intimately  connected as compared to such a close relationship as in relativistic quantum field theory between causality
  and analyticity. In other words, the analyticity
  of the scattering amplitude in nonrelativistic quantum mechanics is not so sacred as  in QFT.
  In relativistic quantum field theories,  the  analyticity property of the scattering amplitude is deeply related with 
  a fundamental principle like microcausality and Lorentz invariance.  We recall  that the nonrelativistic theory   
  is only invariant under Galilean transformations i.e. they are not required to be Poincar\'e invariant. The relativistic quantum field theories
  (QFT) are Poincar\'e invariant. The principle of microcausality plays a crucial role in local field theories.
   Thus the proof of dispersion relations in QFT very critically depends on 
  microcausality. A violation of dispersion relation would necessarily lead to questioning the foundations of general quantum field theories.\\
  (iii) In view of above remarks, we are led to investigate the analyticity property of scattering amplitude in a quantum field theory with
  a compactified spatial dimension.   

\bigskip

\noindent{\bf 2.2a.  Quantum Field Theory with Compact Spatial Dimensions}

\bigskip

\noindent
We have shown in I that the forward scattering amplitude of a theory, defined on the manifold $R^{3,1}\otimes S^1$,  satisfied
dispersion relations. This result was obtained in the frame works of the LSZ formalism. We summarize, in this
subsection, the  starting points of I as stated below.
\\
 We  considered a neutral, scalar field theory with mass $m_0$
 in flat five dimensional Minkowski space $R^{4,1}$. 
It is assumed that the particle is stable and there are no bound states. The notation is  that the  spacetime coordinates
are, $\hat x$,
and all operators are denoted with a {\it hat} when they are defined in the five dimensional  spacetime  where the spatial coordinates 
are noncompact.The  LSZ axioms are \cite{lsz}:\\
{\bf A1.} The states of the system are represented in  a
Hilbert space, ${\hat{ \cal H}}$. All the physical observables are self-adjoint
operators in the Hilbert space, ${\hat{\cal H}}$.\\
{\bf A2.} The theory is invariant under inhomogeneous Lorentz transformations.\\
{\bf A3.} The energy-momentum of the states are defined. It follows from the
requirements of  Lorentz  and translation invariance that
we can construct a representation of the
orthochronous  Lorentz group. The representation
corresponds to unitary operators, ${\hat U}({\hat a},{\hat \Lambda})$,  and the theory is
invariant 
under these transformations. Thus there are Hermitian operators corresponding
to spacetime translations, denoted as ${\hat P}_{{\hat\mu}}$, with ${\hat \mu}=0,1,2,3,4$ which have following
properties:
\be
\bigg[{\hat P}_{\hat\mu}, {\hat P}_{\hat\nu} \bigg]=0
\ee
If ${{\hat{\cal F}}({\hat x})}$ is any Heisenberg operator then its commutator with ${\hat P}_{\hat\mu}$
is
\be
\bigg[{\hat P}_{\hat\mu}, {\hat{\cal F}}({\hat x}) \bigg]=i{\hat\partial}_{\hat\mu} {\hat{\cal F}}({\hat x})
\ee
It is assumed that the operator does not explicitly depend on spacetime 
coordinates.
  If we choose a representation where the translation operators, ${\hat P}_{\hat\mu}$,
are diagonal and the basis vectors $|{\hat p},\hat\alpha>$  span the Hilbert 
space,
${\hat{\cal H}}$, 
\be
{\hat P}_{\hat\mu}|{\hat p},{\hat\alpha}>={\hat p}_{\hat\mu}|{\hat p},\hat\alpha>
\ee
then we are in a position to make more precise statements: \\
${\bullet}$ Existence of the vacuum: there is a unique invariant vacuum state
$|0>$ which has the property
\be
{\hat U}({\hat a},{\hat\Lambda})|0>=|0>
\ee
The vacuum is unique and is Poincar\'e invariant.\\
${\bullet}$ The eigenvalue of ${\hat P}_{\hat\mu}$, ${\hat p}_{\hat\mu}$,   satisfies ${\hat p}_{\hat\mu}{\hat p}^{\hat\mu}>0$ and 
 ${\hat p}_0>0$ for massive theories.
We are concerned  only with  massive stated in this discussion. If we implement
infinitesimal Poincar\'e transformation on the vacuum state then
\be
{\hat P}_{\hat\mu}|0>=0,~~~ {\rm and}~~~ {\hat M}_{\hat{\mu}\hat\nu}|0>=0
\ee
from above postulates and note that ${\hat M}_{{\hat\mu}\hat\nu}$ are the generators of Lorentz
transformations.\\
{\bf A4.} The locality of theory implies that a (bosonic) local operator 
at spacetime point
${\hat x}^{\hat\mu}$ commutes with another (bosonic) 
local operator at ${\hat x}'^{\hat\mu}$ when  their
separation is spacelike i.e. if $({\hat x}-{\hat x}')^2<0$. Our Minkowski metric convention
is as follows: the inner product of two 5-vectors is given by
${\hat x}.{\hat y}={\hat x}^0{\hat y}^0-{\hat x}^1{\hat y}^1-...-{\hat x}^{4}{\hat y}^{4}$.
Since we are dealing with a neutral scalar
field, for the field operator ${\hat\phi }({\hat x})$: ${{\hat\phi}({\hat x})}^{\dagger}={\hat\phi}({\hat x})$ i.e.
 ${\hat\phi} ({\hat x})$ is Hermitian.
By definition it  transforms as a scalar under inhomogeneous Lorentz
transformations 
\be {\hat U}({\hat a},{\hat\Lambda}){\hat\phi}({\hat x}){\hat U}({\hat a},{\hat\Lambda})^{-1}={\hat\phi}({\hat\Lambda} {\hat x}+{\hat a})
\ee
The micro causality, for two local field operators,  is stated to be 
\be
\bigg[{\hat\phi}({\hat x}),{\hat\phi}({\hat x}') \bigg]=0,~~~~~for~~({\hat x}-{\hat x}')^2<0
\ee
It is well known that, in the LSZ formalism,  we are concerned with vacuum
expectation values of time ordered products of operators as well as
with the  the retarded product of fields. The requirements of the above listed axioms
lead to certain relationship, for example, between vacuum expectation values of
R-products of operators. Such a set of relations are termed as the {\it linear relations} and the importance of
the above listed axioms is manifested through these relations. In contrast, unitarity
imposes {\it nonlinear}  constraints on amplitude. For example, if we expand an amplitude
in partial waves, unitarity demands  certain positivity conditions to be satisfied by
the partial wave amplitudes. \\
We summarize below some of the important aspects of LSZ formalism as we utilize them
through out the present investigation. Moreover, the conventions and definitions of I will be followed
for the conveniences of the reader.\\
(i) The asymptotic condition: According to LSZ the field theory accounts for the asymptotic observables.
These correspond to particles of definite mass, charge and spin etc. ${\hat \phi}^{in}({\hat x})$ represents
the free field and a  Fock space is generated by the field operator. The physical observable can be
expressed in terms of these fields.\\
(ii)  ${\hat\phi}( \hat x)$ is the interacting field. LSZ  technique incorporates a prescription to relate the
interacting field, ${\hat\phi}( \hat x)$, with ${\hat\phi}^{in}({\hat x})$; consequently, the asymptotic fields are
defined with a suitable limiting procedure. Thus it is customary to  introduce the notion of the adiabatic switching off
of the interaction. A cutoff adiabatic function is postulated such that this function controls the 
interactions. It is $\bf 1$ at finite interval of time and it has a smooth limit of passing to zero
as $|t| \rightarrow \infty$. It is argued that when adiabatic switching is removed we can define
the physical observables.\\
(iii) The fields  ${\hat \phi}^{in}({\hat x})$ and ${\hat\phi}( \hat x)$ are related as follows:
\be
\label{z}
{\hat x}_0\rightarrow -\infty~~~~{\hat\phi}({\hat x})\rightarrow {\hat Z}^{1/2}{\hat\phi}^{in}({\hat x})
\ee
By the first postulate, ${\hat\phi}^{in}({\hat x})$ creates free particle states. However,
in general ${\hat\phi}({\hat x})$ will create multi particle states besides the single
particle one since it is the interacting field. Moreover, $<1|{\hat\phi}^{in}({\hat x})|0>$ 
and
 $<1|{\hat\phi}({\hat x})|0>$ carry same functional dependence in $\hat x$.  If the factor 
of $\hat Z$ were not the scaling relation between the two
fields (\ref{z}), then canonical commutation relation for each of the 
two fields ( i.e. ${\hat\phi}^{in}({\hat x})$ and  ${\hat\phi}({\hat x})$)  will be the same.
Thus in the absence of $\hat Z$ the two theories will be identical. Moreover, the
postulate of asymptotic condition states that in the remote future
\be
{\hat x}_0\rightarrow \infty~~~~{\hat\phi}({\hat x})\rightarrow {\hat Z}^{1/2}{\hat\phi}^{out}({\hat x}).
\ee
We may as well construct a Fock space utilizing ${\hat\phi}^{out}({\hat x})$ as we could   with ${\hat\phi}({\hat x})^{in}$.
Furthermore, the vacuum is unique for ${\hat\phi}^{in}$,  ${\hat\phi}^{out}$ and ${\hat\phi}({\hat x})$. The
normalizable single particle states are the same i.e.
${\hat\phi}^{in}|0>={\hat\phi}^{out}|0>$. We do not display ${\hat Z}$ from now on. If at all
any need arises,  ${\hat Z}$ can be introduced in the relevant expressions.\\
We define creation and annihilation operators for ${\hat\phi}^{in}$,
${\hat\phi}^{out}$. We recall that   ${\hat\phi}({\hat x})$  is not a free field. Wheheas the fields ${\hat\phi}^{in,out}({\hat x})$ satisfy the free field 
equations $[{\Box}_5+m^2_0]{\hat\phi}^{in,out}({\hat x})=0$, the interacting  field satisfies an equation of motion
which is endowed with  a source current: $[{\Box}_5+m^2_0]{\hat\phi}({\hat x})={\hat j}({\hat x})$.
 We  may use  the plane wave basis for simplicity in certain computations; however,
in a more formal approach, it is desirable to use wave packets.\\
The relevant vacuum expectation values of the products of operators in LSZ formalism are either the time ordered
products (the T-products) or the retarded products (the R-products). We shall mostly use the R-products and 
we use them extensively  throughout this investigation. It is defined as 
\bea
R~{\hat\phi}({\hat x}){\hat\phi}_1({\hat x}_1)...{\hat\phi}_n({\hat x}_n)=&&(-1)^n\sum_P\theta({\hat x}_0-{\hat x}_{10})
\theta({\hat x}_{10}-{\hat x}_{20})...\theta({\hat x}_{n-10}-{\hat x}_{n0})\nonumber\\&&
[[...[{\hat\phi}({\hat x}),{\hat\phi}_{i_1}({\hat x}_{i_1})],{\hat\phi}_{i_2}({\hat x}_{i_2})]..],{\hat\phi}_{i_n}({\hat x}_{i_n})]
\eea
note that $R{\hat\phi}({\hat x})={\hat\phi}({\hat x})$ and  P stands for all the permutations ${i_1,....i_n}$  of  $1,2...n$.
The R-product is hermitial for hermitial fields ${\hat\phi}_i({\hat x}_i)$ and
the product is symmetric under exchange of any fields
${\hat\phi}_1({\hat x}_1)...{\hat\phi}_n({\hat x}_n)$. Notice that the field ${\hat\phi}({\hat x})$ is kept where it is
located in  its position.
We list below some of the important properties of the $R$-product for future use \cite{fr1}:\\
(i) $R~{\hat\phi}({\hat x}){\hat\phi}_1({\hat x}_1)...{\hat\phi}_n({\hat x}_n) \ne 0$ only if
${\hat x}_0>~{\rm max}~\{{\hat x}_{10},..{\hat.x}_{n0} \}$.\\
(ii) Another important property of the R-product is that
\be
R~{\hat\phi}({\hat x}){\hat\phi}_1({\hat x}_1)...{\hat\phi}_n({\hat x}_n) = 0
\ee
whenever the time component ${\hat x}_0$, appearing in the argument of ${\hat\phi}({\hat x})$ whose
position is held fix, is less than time component of any of the  vectors
$({\hat x}_1,...{\hat x}_n)$ appearing in the arguments of ${\hat\phi}({\hat x}_1)...{\hat\phi}({\hat x}_n)$.\\
(iii) We recall that
\be
{\hat\phi}({\hat x}_i)\rightarrow {\hat\phi}({\hat\Lambda} {\hat x}_i)={\hat U}({\hat\Lambda},0){\hat\phi}({\hat x}_i){\hat U}({\hat\Lambda},0)^{-1}
\ee
Under Lorentz transformation ${\hat U}({\hat\Lambda},0)$. Therefore,
\be
R~{\hat\phi}({\hat\Lambda}{\hat x}){\hat\phi}({\hat\Lambda}{\hat x}_i)...{\hat\phi}_n({\hat\Lambda} {\hat x}_n)={\hat U}({\hat\Lambda},0)
R~\phi(x)\phi_1(x_1)...\phi_n(x_n)U(\Lambda,0)^{-1}
\ee
And
\be
 {\hat\phi}_i({\hat x}_i)\rightarrow{\hat\phi}_i({\hat x}_i+{\hat a})=e^{i{\hat a}.{\hat P}}{\hat\phi}_i({\hat x}_i)e^{-i{\hat a}.{\hat P}}
\ee
 under spacetime translations. Consequently,
\be
R~{\hat\phi}( {\hat x}+{\hat a}){\hat\phi}({\hat x}_i+{\hat a})...{\hat\phi}_n({\hat x}_n+{\hat a})=
e^{i{\hat a}.{\hat P}}R~{\hat\phi}({\hat x}){\hat\phi}_1({\hat x}_1)...{\hat\phi}_n({\hat x}_n)e^{-i{\hat a}.{\hat P}}
\ee
 Therefore,   the vacuum expectation value of the R-product
dependents only on  difference between pair of coordinates: in other words it
depends on the
following set of coordinate differences: 
${\hat\xi}_1={\hat x}_1-{\hat x},{\hat\xi}_2={\hat x}_2-{\hat x}_1...{\hat\xi}_n={\hat x}_{n-1} -{\hat x}_n$ as a consequence of
translational invariance. \\
(iv) The retarded property of R-function and the asymptotic conditions lead
 to the following relations.
\bea
[R~{\hat\phi}({\hat x}){\hat\phi}_1({\hat x}_1)...{\hat\phi}_n({\hat x}_n),{\hat\phi}^{in}_l({\hat y}_l)]=&&
i\int d^5{\hat y}'_l\Delta({\hat y}_l-{\hat y'_l})({\Box}_{{5\hat y}'}+{\hat m}_l^2)\times \nonumber\\&&
R~{\hat\phi}({\hat x}){\hat\phi}_1({\hat x}_1)...{\hat\phi}_n({\hat x}_n){\hat\phi}_l({\hat y}'_l)
\eea
Note: here ${\hat m}_l$ stands for the mass of a field in five dimensions.
We may define 'in' and 'out' states in terms of the creation operators associated with 'in' and 'out' fields as follows
\be
\label{fock1}
|{\hat k}_1,{\hat k}_2,....{\hat k}_n~in>={\hat a}_{in}^{\dagger}({\hat{\bf k}}_1){\hat a}_{in}^{\dagger}({\hat{\bf k}}_2)...
{\hat a}_{in}^{\dagger}({\hat{\bf k}}_n)|0>
\ee
\be
\label{fock2}
|{\hat k}_1,{\hat k}_2,....{\hat k}_n~out>={\hat a}_{out}^{\dagger}({\hat{\bf k}}_1){\hat a}_{out}^{\dagger}({\hat{\bf k}}_2)...
{\hat a}_{out}^{\dagger}({\hat{\bf k}}_n)|0>
\ee
We can construct a complete set of states either starting from 'in'  field operators or the 'out' field operators and each complete set
will span the Hilbert space,  ${\hat{\cal H}}$. Therefore, a unitary operator will relate the two sets of states in this Hilbert
space. This is a heuristic way of introducing the concept of the $S$-matrix. We shall define $S$-matrix elements
through LSZ reduction technique in subsequent section.\\
We shall not distinguish between notations like ${\hat\phi}^{out,in}$ or ${\hat\phi}_{out,in}$ and therefore, there might be use
of the sloppy notation in this regard.\\
We record the following important remark {\it en passant}. The generic matrix element 
$<{\hat\alpha}|{\hat\phi}({\hat x}_1){\hat\phi}({\hat x}_2)...|{\hat\beta}>$
is not an ordinary function but a distribution. Thus it is to be always
understood as smeared with a Schwarz type test function $f\in {\cal S}$. The test
function is infinitely differentiable and it goes to zero along with all its
derivatives faster than any power of its argument. We shall formally derive expressions
for scattering amplitudes and the absorptive parts by employing the LSZ technique. It is to be understood that
these are generalized functions and such matrix elements are properly defined
with smeared out test functions.\\
We  obtain below the expression for the K\"allen-Lehmann representation for the
five dimensional theory. It will help us to transparently expose, as we shall recall in the next section, the consequences of
$S^1$ compactification. Let  us consider the vacuum expectation value (VEV) of the
commutator of two fields in the $D=5$ theory: $<0|[{\hat\phi}({\hat x}), {\hat\phi}({\hat y})]|0>$. We
introduce a complete set of states between product of the fields after opening up the commutator. Thus
we arrive at the following expression by adopting the standard arguments,
\bea
\label{KL}
  <0|[{\hat\phi}({\hat x}), {\hat\phi}({\hat y})]|0>=\sum_{\hat\alpha}\bigg(<0|{\hat\phi}(0){\hat\alpha}>e^{-i{\hat p}_{\hat\alpha}.({\hat x}-{\hat y})}
<{\hat\alpha}|{\hat\phi}(0)|0>-({\hat x}{\leftrightarrow}{\hat y}) \bigg)
\eea
Let us define
\be
{\hat\rho}({\hat q})=(2\pi)^4\sum_{\hat\alpha}\delta^5({\hat q}-{\hat p}_{\hat\alpha})|<0|{\hat\phi}(0)|{\hat\alpha}>|^2
\ee
Note that ${\hat\rho}({\hat q})$ is positive, and ${\hat\rho}=0$ when ${\hat q}$ is not in the light cone. It is also Lorentz
invariant. Thus we write
\be
{\hat\rho}({\hat q})={\hat\sigma}({\hat q}^2)\theta({\hat q}_0),~~{\hat\sigma}({\hat q}^2)=0,~~~if~~{\hat q}^2<0
\ee
This is a positive measure. We may separate the expression for the VEV of the commutator (\ref{KL}) into two parts:
the single particle state contribution and the rest. Moreover, we use the asymptotic state condition to arrive at
\bea
\label{KL1}
<0|[{\hat\phi}(\hat x}),{\hat\phi}({\hat y})]|0>=i{\hat Z}{\hat\Delta}({\hat x},{\hat y}; m_0)+i\int_{{\hat m}_1^2}^{\infty}d{\hat m}'^2{\hat{\Delta}
({\hat x},{\hat y}; {\hat m}')
\eea
where ${\hat\Delta}({\hat x},{\hat y};m_0)$ is the VEV of the free field commutator, $m_0$ is the mass of the scalar. ${\hat m}_1^2>{\hat M}^2$, the multiple
particle threshold.\\
We are in a position to  study several attributes of scattering amplitudes in the five dimensional theory such as proving existence of
the Lehmann-Martin ellipse, give a proof of fixed t dispersion relation to mention a few. However, these properties
have been derived in a general setting recently  by us \cite{jmjmp1} for D-dimensional theories.
The purpose of incorporating the expression for the VEV of the commutator of two fields in the 5-dimensional theory
is to provide a prelude to the modification of similar expressions when we compactify the theory on $S^1$ as we shall
see in the next section.

\bigskip

\noindent{\bf 2.2b. Compactification of Scalar Field Theory: } ${\bf R^{4,1} \rightarrow  R^{3,1}\otimes S^1}$

\bigskip

\noindent  In this subsection,  $S^1$ compactification of a spatial coordinate of the five dimensional theory is considered. To start with, decompose the
five dimensional spacetime coordinates, ${\hat x}^{\hat\mu}$,  as follows:
\be
 {\hat x}^{\hat\mu}=(x^{\mu}, y)
  \ee
  where $x^{\mu}$ are the four dimensional Minkowski  space coordinates; $y$ is the compact coordinate on $S^1$ with periodicity
   $y+2\pi R = y$, $R$ being the radius of $S^1$. We summarize below  the attributes this $S^1$
  compactification. The neutral scalar field of mass $m_0$ defined  in $D=5$  manifold  is  now described in the geometry $R^{3,1}\otimes S^1$. 
  We focus on the  free field version {\it such as the  'in' and 'out' field} ,  ${\hat\phi}^{in, out}({\hat x})$. The equation of motion is
  $[{\Box}_5+m_0^2]{\hat\phi}^{in,out}({\hat x})=0$. We expand the field 
  \be
  \label{kk1}
  {\hat\phi}^{in,out}({\hat x})={\hat\phi}^{in,out}(x,y)=\phi^{in,out}_0(x)+
\sum_{n=-\infty, n\ne 0}^{+n=\infty}\phi^{in,out}_n(x)e^{{{in y}\over{ R}}}
  \ee
  Note that $\phi^{in,out}_0(x)$, the so called zero mode, has no $y$-dependence. The terms in rest of the series  (\ref{kk1})
  satisfy periodicity in $y$. The  five dimensional Laplacian, ${\Box}_5$, is decomposed  as sum two operators: 
    $\Box_4$ and  ${{\partial}\over{\partial y^2}}$ . The equation of motion is
  \be
  \label{kk2}
  [\Box_4 - {{\partial}\over{\partial y^2}}+m_n^2]\phi^{in,out}_n(x,y)=0
  \ee
  where $\phi^{in,out}_n(x,y)=\phi_n^{in,out}e^{{{in y}\over{ R}}}$ and $n=0$ term has no $y$-dependence being $\phi_0(x)$; from now on
  $\Box_4=\Box$.
  Here $m_n^2=m_o^2+{{n^2}\over{R^2}}$. Thus we have tower of massive states. The momentum associated
  in the $y$-direction is $q_n=n/R$ and is quantized  in the units of $1/R$. It is an additive conserved quantum
  number. We term it as Kaluza-Klein (KK) charge although there is no gravitational interaction in the five dimensional theory
  we started with and 
   still call it KK reduction.  For the interacting field ${\hat\phi}({\hat x})$, we can adopt a similar mode expansion. 
   \be
   \label{kk2x}
   {\hat\phi}({\hat x})={\hat\phi}(x,y)=\phi_0(x)+\sum_{n=-\infty, n\ne 0}^{n=+\infty}\phi_n(x)e^{{{iny}\over R}}
   \ee
   The equation of motion for the interacting fields is endowed with a source term. Thus source current would be expanded
   as is the expansion (\ref{kk2x}). Each field $\phi_n(x)$ will have a current, $J_n(x)$ associated with it and source
   current will be expanded as
   \be
   \label{kk2a}
   {\hat j}(x,y) =j_0(x)+\sum_{n=-\infty, n\ne 0}^{n=+\infty}J_n(x)e^{i{ny/R}}
   \ee
   Note that the set of currents,  $\{J_n(x)  \}$, are the source currents associated with the tower of interacting fields
   $\{ \phi_n(x) \}$. These fields carry the discrete KK charge, $n$. Therefore, $J_n(x)$ also carries the
   same KK charge. We should keep this aspect in mind when we consider matrix element of such currents between
   stated. In future, we might not explicitly display the charge of the current; however, it becomes quite obvious
   in the context.\\ 
  The zero mode, $\phi^{in,out}_0$, create their  Fock spaces. Similarly, each of the fields $\phi^{in,out}_n(x)$ create  their Fock spaces as well.
  For example a state with spatial momentum, ${\bf p}$, energy, $p_0$ and discrete momentum $q_n$ (in $y$-direction) is created
  by
  \be
   \label{kk3}
   A^{\dagger}({\bf p},q_n)|0>=|p,q_n>,~~p_0>0 
   \ee
  {\it Ramark:} The five dimensional has a neutral, massive scalar field. theory. 
  After the $S^1$ compactification to  the $R^{3,1}\otimes S^1$, the spectrum of the resulting theory consists of 
  a massive field of mass $m_0$, associated with the zero mode and tower of Kaluza-Klein (KK) states  characterized by a mass and a 'charge', $(m_n, q_n)$,
  respectively. We can discuss the structure of the Hilbert space of the compactified theory. \\ 
   {\it The Decomposition of the Hilbert space ${\hat{\cal H}}$:} The Hilbert space associated with the five dimensional theory is ${\hat {\cal H}}$. It is now decomposed as a direct sum of Hilbert spaces where each one is characterized by its quantum number $q_n$   
  \be
  \label{kk4}
  {\hat{\cal H}}=\sum \oplus {\cal H}_n
  \ee
  Thus ${\cal H}_0$ is the Hilbert space constructed from $\phi_0^{in,out}$ with charge $q_{n=0}$. This space is built by the actions of the creation operators 
  $\{ a^{\dagger}({\bf k}) \}$  acting
  on the vacuum and these states span ${\cal  H}_0$. 
   A single particle state is  $a^{\dagger}({\bf k})|0>=|{\bf k}>$ and multiparticle states
  are created using the procedure out lines in (\ref{fock1}) and (\ref{fock2}). We can create Fock spaces by the actions of fields
   $\phi_n(x,y)$ with  charge $q_n$ on the vacuum. This space is constructed through the action of creation operators
   $\{A^{\dagger}({\bf p}, q_n)\}$. Now  two  state vectors with different 'charges'
  are orthogonal to one another 
  \be
  \label{kk5}
  <{\bf p}, q_{n'}|{\bf p}',q_{n'}>=\delta^3({\bf p}-{\bf p'})\delta_{n,n'}
   \ee
  {\it Remark: } We assume that there are no bound states in the theory and all particles
  are stable as mentioned. There exists a possibility that a particle with charge $2n$ and mass $m^2_{2n}=m_0^2+{{4n^2}\over{R^2}}$ could be a
  bound state of two particles of charge $n$ and masses $m_n$ each under certain circumstances. We have excluded such possibilities
  from the present investigation. \\
  The LSZ formalism can be adopted for the compactified theory. If we keep in mind the steps
  introduced above, it is possible to envisage field operators $\phi^{in}_n(x)$ and $\phi^{out}_n(x)$ for each of the fields for a given $n$.
  Therefore, each Hilbert space, ${\cal H}_n$ will be spanned by the state vectors created by operators $a^{\dagger}({\bf k})$, for $n=0$ and
  $A^{\dagger}({\bf p}, q_n)$, for $n\ne 0$.  Moreover, we are in a position to define corresponding set of interacting field $\{\phi_n(x) \}$ which
  will interpolate  into 'in' and 'out' fields in the asymptotic limits.\\
  {\it Remark}: Note that in (\ref{kk1}) sum over $\{n\}$ runs over positive and negative integers. If there is a parity symmetry $y\rightarrow -y$
  under which the field is invariant we can reduce the sum to positive $n$ only. However, since $q_n$ is an additive discrete quantum number, 
  a state with $q_n>0$ could be designated as a particle and the corresponding state $q_n<0$ can be interpreted as its antiparticle. Thus
  a two particle state $|p,q_n>|p,-q_n>,~~q_n>0~and ~p_0>0$ is a particle 
antiparticle state, $q_n=0$; in other words the sum of the
total charges of the two states is zero.  Thus it has the quantum
number of the vacuum. For example,
  it could be two particle state of $\phi_0$ satisfying energy momentum conservation, especially if they appear as intermediate states. 
   \\
   Now return to  the K\"allen-Lehmann representation (\ref{KL}) in the present context and utilize the expansion (\ref{kk2x}) in the 
   expression for the VEV of the commutator of two fields defined in $D=5$:  $<0|[{\hat\phi}({\hat x}),{\hat\phi}({\hat x}')]|0>$
   \bea
   \label{kk6}
  <0|[{\hat\phi}(x,y),{\hat\phi}(x',y')]|0>=<0|[\phi_0(x)+\sum_{-\infty}^{+\infty}\phi_n(x,y),~ \phi_0(x')+\sum_{-\infty}^{+\infty}\phi_l(x',y')]|0> 
   \eea
   The VEV of a commutator of two fields  given by the  spectral representation (\ref{KL}) will be decomposed into sum of several commutators
   whose VEV will appear:
   \bea
   \label{kk7}
   <0|[\phi_0(x), \phi_0(x')]|0>,~~<0|[\phi_n(x),\phi_{-n}(x')]|0>,...
   \eea
   Since the vacuum carries zero KK charge, $q_{vac}=0=q_0$, the commutator of 
two fields (with $n\ne 0$) should give rise to zero-charge and 
   only $\phi_n$ and $\phi_{-n}$ commutators will appear.
    Moreover, commutator of fields with different $q_n$ vanish since the operators act on states of different Hilbert spaces. Thus
   we already note the consequences of compactification. When we wish to evaluate the VEV and insert complete set of intermediate
   states in the product of two operators after opening up the commutators, we note that all states of the entire KK tower can appear
   as intermediate states as long as they respect all conservation laws. This will be an important feature in all our computations in
   what follows.

   \bigskip
   
   \noindent{\bf{2.3.  Definitions and Kinematical Variables }}
   
   \bigskip
   
   \noindent  The purpose of this investigation is to derive analyticity property of the fixed-$t$ dispersion relations for scattering
   of the KK states carrying nonzero charge i.e. scattering in the $q_n\neq 0$ sector. However, we mention in passing the other
   possible processes. These are (i) scattering of states with $q_n=0$ states, i.e. scattering of zero modes. (ii)
   The scattering of a state carrying charge $q_n=0$ with a state with non-zero KK charge. We have studied 
   reactions (i) and (ii) in I and therefore, we do not wish to dwell upon them here.\\ 
   We shall define the kinematical variables below. The states carrying $q_n\ne 0$ are denoted by $\chi_n$ (from now on a state
   carrying charge is defined with a subscript $n$ and momenta carried by external particles are denoted as $p_a, p_b,...$. Moreover, we shall consider elastic scattering of states carrying equal charge;
   the elastic scattering of unequal charge particles is just elastic scattering of unequal mass states due to mass-charge relationdhip
   for the KK states.\\   
Let us consider a generic 4-body reaction (all states carry non-zero $n$)
\be
\label{kk6}
 a + b\rightarrow c + d 
\ee
The particles $( a,  b,  c,  d) $ (the corresponding fields being $\chi_a, \chi_b, \chi_c, \chi_d$)
respectively carrying momenta $ {\tilde p}_a, {\tilde p}_b, {\tilde p}_c, {\tilde p}_d$; these particles may
correspond to the KK zero modes (with KK momentum $q=0$) or particles might carry nonzero KK charge. We shall consider
only elastic scatterings. The Lorentz invariant Mandelstam variables are
\be
\label{kk7}
s=( p_a+ p_b)^2=( p_c+ p_d)^2,~t=( p_a- p_d)^2=( p_b- p_c)^2,~
u=( p_a-p_c)^2=( p_b- p_d)^2
\ee
and $\sum  p^2_a+ p^2_b+ p^2_c+ p^2_d=m_a^2+m_b^2+m_c^2+m_d^2$.
The independent identities of the four particles will facilitate the computation amplitude so that to keep track
of the fields reduced using LSZ procedure.  
  We list below some relevant (kinematic) variables which will be required in future
\be
\label{kk8}
{\bf M}_a^2,~~{\bf M}_b^2,~~{\bf M}_c^2,~~{\bf M}_d^2
\ee
These correspond to lowest mass two or more particle states which carry the same quantum number as that of
particle $a$, $b$, $c$ and $d$ respectively. We  define below six more variables 
\be
\label{kk9}
({\bf M}_{ab}, {\bf M}_{cd)},~~({\bf M}_{ac}, {\bf M}_{bd}),~~({\bf M}_{ad}, {\bf M}_{bc})
\ee
The variable ${\bf M}_{ab}$ carries the same quantum number as $(a ~and~ b)$ and it corresponds to two or more particle
states. Similar definition holds for the other five variables introduced above. 
 We define two types of thresholds: (i) the physical threshold, $s_{phys}$, and $s_{thr}$. In absence of anomalous thresholds (and equal mass scattering)
 $s_{thr}=s_{phys}$. Similarly, we may define $u_{phys}$ and $u_{thr}$ which will be useful when we discuss dispersion relations.  We assume from 
 now on that $s_{thr}=s_{phys}$ and $u_{thr}=u_{phys}$.
Now we outline the derivation of the expression a four point function  in the  LSZ formalism.  We  start with 
$| p_d, p_c~out>$ and $| p_b, p_a~in>$ and considers the matrix element $< p_d, p_c~out|  p_b, p_a~in>$.
Next we subtract out the matrix element $< p_d, p_c~in| p_b, p_a~in>$ to define the S-matrix element.
\bea
\label{kk10}
< p_d, p_d~out| p_b,p_a~in>=&&\delta^3({\bf p}_d-{\bf p}_b)\delta^3({\bf p}_c-{\bf p}_a)
-{{i}\over{(2\pi)^3}} \int d^4x\int d^4x' \nonumber\\&&e^{-i( p_a.x- p_cx')}K_xK_{x'}< p_d~out|R(x',x)| p_b~in>
\eea
where $K_x$ and $K_{x'}$ are the four dimensional Klein-Gordon operators and
\be
\label{kk11}
R(x,x')=-i\theta (x_0-x_o')[\chi_a(x),\chi_c(x')]
\ee
We have reduced fields associated with $a$ and $c$ in (\ref{kk10}). 
In the next step we may reduce all the four fields and in such a reduction we shall get VEV of the R-product of four fields which will be operated
upon by four K-G operators. However, the latter form of  LSZ reduction (when all fields are reduced) 
is not very useful when we want to investigate the analyticity property of the
amplitude in the present context.
 In particular our intent is  to write the forward dispersion relation. Thus we abandon the idea of reducing all the four fields.\\
{\it Remark: } Note that on the right hand side of the requation (\ref{kk10}) the operators act on
 $R\chi_a(x)\chi(x')_c$ and there is a $\theta$-function
in the definition of the R-product. Consequently, the action of $K_xK_{x'}$ on $R\chi_a(x)\chi_c(x')$ will produce a term 
$R J_a(x) J_c(x')$. In addition the operation of the two K-G operators will give rise to $\delta$-functions and derivatives of $\delta$-functions
and some equal time commutators i.e. there will terms whose coefficients are $\delta (x_0-x_0')$. When we consider fourier transforms of 
the derivatives of these $\delta$-function derivative terms they will be transformed to momentum variables. However, the amplitude is a function of
Lorentz invariant quantities. Thus one will get only finite polynomials of such variables, as has been argued by Symanzik \cite{kurt}.
His arguments is that  in a local quantum field
theory only finite number of derivatives of $\delta$-functions can appear. Moreover, in addition, there are some 
equal time commutators and many of them vanish when we
invoke locality arguments. Therefore, we shall use the relation
\be
\label{kk12}
K_xK_{x'}R\chi(x)\chi_c(x')=R J_a(x) J_c(x')
\ee
 keeping in mind that there are derivatives of $\delta$-functions and some equal time commutation relations which might be present.
 Moreover, since the derivative terms give rise to polynomials in Lorentz invariant variables, the analyticity properties of the amplitude
 are not affected due to the presence of such terms. This will be understood whenever we write an equation like (\ref{kk12}). 

  \bigskip
  
  \noindent {\bf 3.  The Nonforward Elastic Scatting  of $n \ne 0$ Kaluza-Klein States}
  
  \bigskip
  
  \noindent  We envisage elastic scattering of two equal mass, $m_n^2=m_0^2+{{n^2}\over{R^2}}$,  hence equal charge KK particles 
  and we take $n$ positive. 
  Our first step is to define the scattering amplitude for this reaction (see \ref{kk10}) 
\bea
 \label{nn1}
 <p_d,p_c~out|p_b,p_a~in>=&& 4p^0_ap^0_b\delta^3({\bf p}_d-{\bf p}_b)\delta^3({\bf p}_a-{\bf p}_c) -\nonumber\\&& {{i}\over{(2\pi)^3}}
 \int d^4x\int d^4x'e^{-i(p_a.x-p_c.x')}\times \nonumber\\&& {\tilde K}_x{\tilde K}_{x'}<p_d~out|{\bar R}(x';x)|p_b~in>
 \eea
  where
\be
 \label{nn2}
 {\bar R}(x';x)=-i\theta (x_0-x_0')[\chi_a(x),\chi_c(x')]
 \ee  
  and ${\tilde K}_x=(\Box+m_n^2)$. We let the two KG operators act on ${\bar R}(x;x')$  in the VEV and resulting
  equation is
  \bea
  \label{nn3}
 <p_d,p_c~out|p_b,p_a~in>=&&<p_d,p_c~in|p_b,p_a~in> 
  -{{1}\over{(2\pi)^3}}
 \int d^4x\int d^4x'e^{-i(p_a.x-p_c.x')} \times \nonumber\\&&
  <p_d|\theta(x_0'-x_0)[J_c(x'),J_a(x)]|p_b>
  \eea
  Here $J_a(x)$ and $J_c(x')$ are the source currents associated with the fields $\chi_a(x)$ and $\chi_b(x')$ respectively.
  We arrive at (\ref{nn3}) from (\ref{nn1}) with the understanding that the $R.H.S.$ of (\ref{nn3}) contains additional
  terms; however, these terms do not affect the study of the analyticity properties of the amplitude as alluded to earlier.
We shall define three distributions which are matrix elements of the product of current. The importance of these
functions will be evident in sequel   
\bea
 \label{nn4}
    F_R(q)=\int_{\infty}^{+\infty}d^4ze^{iq.z}\theta(z_0)<Q_f|[J_a(z/2),J_c(-z/2)]|Q_i>
    \eea
 \bea
    \label{nn5}
 F_A(q)=-\int_{\infty}^{+\infty}d^4ze^{iq.z}\theta(-z_0)<Q_f|J_a(z/2),J_c(-z/2)]|Q_i>
 \eea
 and
 \be
 \label{nn6}
 F_C(q)= \int_{-\infty}^{+\infty}d^4ze^{iq.z}<Q_f|[J_a(z/2),J_c(-z/2)]|Q_i>
 \ee
Moreover,
\be
\label{nn7}
F_C(q)=F_R(q)-FA(q)
\ee 
  $|Q_i>$ and $|Q_f>$ are states which carry four momenta and these momenta are held fixed. At this stage we treat them as
  parameter;  
 it is elaborated  in ensuing discussions. Let us focus attention on the matrix element of the causal commutator defined in (\ref{nn6}). We open
 up the commutator of the currents and introduce a complete set of physical states. Let us assign KK charge $n$ to each of
 the states. Thus the conservation of KK charge only permits those intermediate states which respect the charge
 conservation laws.  The physical complete sets are:   $\sum_n|{\cal P}_n{\tilde\alpha}_n><{\cal P}_n{\tilde\alpha}_n|={\bf 1}$ and
 $\sum_{n'}|{\bar{\cal P}}_{n'}{\tilde\beta}_{n'}><{\bar{\cal P}}_{n'}{\tilde\beta}_{n'}|={\bf 1}$.
Here $\{{\tilde\alpha}_n, {\tilde\beta}_{n'} \}$
stand for quantum numbers that are permitted for the intermediate states. 
The matrix element defining $F_C(q)$, (\ref{nn7}), assumes the following form
 \bea
\label{nn8}
&&\int d^4ze^{iq.z}\bigg[\sum_n\bigg(\int d^4{\cal P}_n<Q_f|J_a({z\over 2})|{\cal P}_n{\tilde\alpha}_n
><{\cal P}_n{\tilde\alpha}_n|J_c(-{z\over 2})|Q_i>\bigg)\nonumber\\&& -
\sum_{n'}\bigg(\int d^4{\bar{\cal P}}_{n'}<Q_f|J_c(-{z\over 2})|{\bar{\cal P}}_{n'}{\tilde\beta}_{n'}>
<{\bar{\cal P}}_{n'}{\tilde\beta}_{n'}|J_a({z\over 2})|
Q_i>\bigg) \bigg]
\eea
We proceed as follows at this point. Let us use translation operations judiciously so that the currents do not carry
any dependence in the $z$-variables.  Subsequently, we integrate over $d^4z$ which leads to $\delta$-functions. 
\newpage
The expression for $F_C(q)$  now takes the form
 \bea
\label{nn9}
&& F_C(q)=\sum_n\bigg(<Q_f|j_a(0)|{\cal P}_n={{(Q_i+Q_f)}\over 2}-q,{\tilde\alpha}_n>\times \nonumber\\&& 
<{\tilde\alpha}_n,{\cal P}_n={{(Q_i+Q_f)}\over 2}-q|j_c(0)|Q_i>\bigg) \nonumber\\&&
-\sum_{n'}\bigg(<Q_f|j_c(0)|{\bar{\cal  P}}_{n'}={{(Q_i+Q_f)}\over 2}+q,{\tilde\beta}_{n'}>\times \nonumber\\&& 
<{\tilde\beta}_{n'},{\bar{\cal P}}_{n'}={{(Q_i+Q_f)}\over 2}+q|j_a(0)|Q_i>\bigg)
\eea
A few explanatory comments are in order: The momentum of the intermediate state ${\cal P}_n$ appearing
in  first term in (\ref{nn9}) is constrained to ${\cal P}_n=({{Q_i+Q_f}\over 2})-q$ after the $d^4z$ integration. Similarly,
${\cal P}_{n'}=({{Q_i+Q_f}\over 2})+q $ in the second term of (\ref{nn9}). The second point is that, in the derivation of
the spectral representation line (\ref{nn9}) for a theory with single scalar field, the physical intermediate states
correspond to the multiparticle states consistent with energy momentum conservation (physical states). For
the case at hand, the intermediate states consist of the entire KK tower as long as these states satisfy
energy momentum conservation constraints and the KK charge conservation rules. We shall discuss the consequences
of this aspect in sequel.   \\
Let us define
 \bea
\label{nn10}
2A_s(q)=&&\sum_{n'}\bigg(<Q_f|j(0)_a|{\bar{\cal P}}_{n'}={{(Q_i+Q_f)}\over 2}+q,{\tilde\beta}_{n'}>
\times \nonumber\\&&
<{\tilde\beta}_n',{\bar{\tilde P}}_n={{(Q_i+Q_f)}\over 2}+q|j_c(0)|Q_i>\bigg)
\eea
and 
\bea
\label{nn11}
2A_u=&&\sum_n\bigg(<Q_f|j_c(0)|{\cal P}_n={{(Q_i+Q_f)}\over 2}-q,{\tilde\alpha}_n>\times
\nonumber\\&&
<{\tilde\alpha}_n,{\cal P}_n={{(Q_i+Q_f)}\over 2}-q|j_l(0)|Q_i>\bigg)
\eea
 {\it Consequences of microcausality:} The Fourier transform of $F_C(q), {\bar{F}}_C(z)$, vanishes outside the light cone.
We recall that, 
\be
\label{nn12}
F_C(q)= {{1}\over 2}(A_u(q)-A_s(q))
\ee
Moreover, $F_C(q)$ will also vanish as function of $q$ wherever, both $A_s(q)$ and $A_u(q)$ vanish simultaneously.
We recall that the the intermediate states are physical states and their four momenta lie in  the
forward light cone, $V^+$, as a consequence
\be
\label{nn13}
({{Q_i+Q_f}\over 2}+q)^2\ge 0,~~~({{Q_i+Q_f}\over 2})_0+q_0\ge 0
\ee
and
\be
\label{nn14}
({{Q_i+Q_f}\over 2}-q)^2\ge 0,~~~({{Q_i+Q_f}\over 2})_0-q_0\ge 0
\ee
The above two conditions, for nonvanishing of $A_u(q)$ and $A_s(q)$ implies existence of minimum mass parameters\\ 
(i) $ ({{Q_i+Q_f}\over 2}+q)^2\ge {{\cal M}_+}^2$ 
and (ii) $({{Q_i+Q_f}\over 2}-q)^2\ge {{\cal M}_-}^2 $.\\
 The matrix elements for
$A_s(q)$ and $A_u(q)$ will not vanish and if the two conditions stated above,
pertinent to each of them, are fulfilled.\\
We would like to draw the attentions of the reader to the following facts in the context a theory with compactified spatial dimension.
In the case where there is only one scalar field, the sum over intermediate physical states as given in (\ref{nn10}) and (\ref{nn11}) is the
multiparticles states  permitted by energy momentum conservations. However, in the present situation, the contributions to 
the intermediate states are those which come from the KK towers as allowed by the charge conservation rules (depending on what
charges we assign to $|Q_i>$ and $Q_f>$ for the elastic scattering) and energy momentum conservation. For example, if the initial
states have change $n=1$, then the tower of multiple particle intermediate states should have one unit of KK charge. Thus the question
is whether the infinity tower of KK states would contribute? It looks like that at the present stage, when we are in the 'linear programme"
framework of the general field theoretic formalism, this issue cannot be resolved. As we shall discuss subsequently, when unitarity
constraint is invoked there are only contributions from finite number of terms as long as $s$ is finite but can be taken to be very large.\\
In order to derive a fixed-$t$ dispersion relation we have to identify a domain which is free from singularities in the $t$-plane. The first step
is to obtain the Jost-lehmann-Dyson representation for the causal commutator, $F_C(q)$.  We are considering elastic scattering of equal mass particles
i.e. all particles carry same KK charge. Therefore, the technique of Jost and Lehmann \cite{jl} is quite adequate; we do not have to resort to
more elegant and general approach of Dyson \cite{dyson} (see  \cite{jmjmp1} for detail discussions). We shall adhere to notations and discussions of
reference I and present those results in a concise manner.
 As noted in (\ref{nn13}) and (\ref{nn14}), $F_C(q)$ is nonvanishing in those domains. We designate this region
as ${\bar{\bf R}}$,
\be
\label{nn15}
{\bar{\bf R}}: \bigg\{(Q+q)^2\ge{{\cal M_+}}^2, Q+q\in V^+ ~ {\rm and} ~ (Q-q)^2 \ge{{\cal M_-}}^2, Q-q\in V^+  \bigg\}
\ee
where $Q={{Q_i+Q_f}\over 2}$ and  $V^+$ being the future light cone. We need not repeat derivation of the Jost-Lehmann representation here. The
present case differs from the case where only one field is present in the following way. Here we are looking for the nearest singularity to
determine the singularity free region. For the case at hand, the presence of the towers of KK states is to be envisaged in 
the following perspective. Since we consider equal mass scattering the location of nearest singularity will be decided by the lowest
values of ${\cal M}_+$ and ${\cal M}_-$. Let us elaborate this point. We recall that there is the tower of KK states appearing as intermediate
states (see (\ref{nn10} ) and (\ref{nn11} )). Thus each new threshold could create region of singularity of $F_C(q)$. We are concerned
about the  identification of the singularity free domain. Thus the lowest threshold of two particle intermediate state, consistent
with desired constraints, control the determination of this domain of analyticity. Therefore,
for the equal mass case, the Jost-Lehmann representation for $F_C(q)$ is such that it is nonzero in the region ${\bar{\bf R}}$,
\be
\label{nn16}
F_C(q)=\int_Sd^4u\int_0^{\infty}d\chi^2\epsilon(q_0-u_0)\delta[(q-u)^2-\chi^2)]
\Phi(u,Q.\chi^2)
\ee
Note that $u$ is also a 4-dimensional vector ({\it not the Mandelstam
variable u}). The domain of integration of $u$ is the region $S$ specified
below
\bea
\label{nn17}
{\bf S}:\bigg\{Q+u \in V^+,~ Q-u \in V^+,~
Max~ [0,{\cal M}_+-\sqrt{(Q+u)^2},{\cal M}_--\sqrt{(Q-u)^2}]\le \chi \bigg\}
\eea
and $ \Phi(u,Q.\chi^2)$ arbitrary.
 Here $\chi^2$ is to be interpreted like a mass parameter. Moreover, recall that
the assumptions about the features of the causal function stated above are
the properties we have listed earlier  and $Q$ is already defined above. Since the retarded
commutator involves a $\theta$-function, if we use integral representation for it (see \cite{jl}) we
derive an expression for the regarded function,
\be
\label{nn18}
F_R(q)={{i\over {2\pi}}}\int d^4q'\delta^3({\bf q'}-{\bf q})
{{1\over{q_0'-q_0}}}F_C(q'), ~Im~q_0>0
\ee
Moreover, for the retarded function, $F_R(q)$, 
the corresponding Jost-Lehmann representation  reads \cite{jl}
\be
\label{nn19}
F_R(q)={{i\over{2\pi}}}\int_Sd^4u\int_0^{\infty}d\chi^2
{{\Phi(u,Q,\chi^2)}\over{(q-u)^2-\chi^2}}
\ee
We mention in passing that these integral representations are written under the assumption that the functions appearing
inside the integral are such that the integral converges. However, if there are polynomial growths asymptotically then
subtraction procedure can be invoked to tame the divergences. It is to be borne in mind that these expressions can have
only polynomial behaviors for asymptotic values of the argument as we have argued earlier. The polynomial behaviors will not affect the study of analyticity
properties. One important observation is that that the singularities lie in the complex $q$-plane \footnote{see Itzykson and Zubber  \cite{book3} and
Sommer \cite{sommer} for elaborate discussions}. We provide  below a short and transparent discussion for the sake of completeness.
The locations of the singularities are found by examining where the denominator (\ref{nn19}) vanishes,
\be
\label{nn20}
(q_0-u_0)^2-(q_1-u_1)^2-(q_2-u_2)^2-(q_3-u_3)^2=\chi^2
 \ee
 We conclude that the the singularities lie on the hyperboloid give by (\ref{nn20}) and those points are in domain $\bf S$ as defined in  
(\ref{nn17}). There are points in the hyperboloid which belong to the domain $\bf S$. These are called admissible regions. Moreover, according our
earlier definition the domain ${\bar{\bf R}}$ is where $F_C(q)$ is nonvanishing (see (\ref{nn15})). Then there  is a domain which contains a set of real 
points where $F_C(q)$ vansishes, call it $\bf R$ and this is compliment to real elements of ${\bar{\bf R}}$. From the above arguments, we arrive
at the conclusion that $F_C(q)=0$  for every real point belonging to ${\bf R}$ (the compliment of ${\bar{\bf R}}$). Thus these are the
real points in the $q$-plane where $F_R(q)=F_A(q)$ since $F_C(q)=0$ there. Recall the definition of ${\bar{\bf R}}$, (\ref{nn15}). A border
is defined by the upper branch of the parabola given by the equation $(Q+q)^2={\cal M_+}^2$ and the other one is given by the equation
for another parabola $(Q-q)^2={\cal M_-}^2$. Now we identify the {\it coincidence region} to be the domain bordered by the two parabolae.
It is obvious from the above discussions  that the set $\bf S$ is defined by the range of values $u$ and $\chi^2$ assume in the admissible
parabola. Now we see that those set of values belong to a subset of $(u,\chi^2)$ of all parabolas (recall equation (\ref{nn20}))  
\cite{sommer} and  \cite{jl,dyson} .  In order to transparently discuss the location of a singularity, let us go through a few short steps as
 the prescription to illustrate essential points.  We discussed about the identification of admissible parabola. The amplitude is function of
Lorentz invariant kinematical variables; therefore, it is desirable to express the constraints and equations in terms of those variables
eventually. Let us focus on $Q\in V^+$ and choose a Lorentz frame such that four vector $Q=(Q_0,{\bf 0})$ where $\bf 0$ stands for the
{\it three} spatial components of $Q$. Next step is to choose four vector $q$ appropriately to exhibit the location of singularity in a simple way.
This is achieved as follows: choose one spatial component of $q$ in order to identify the position of the singularity in this variable and
treat $q_0$ and the rest of the components of $q$ as parameters and hold them fixed \cite{sommer}. We remind the reader that all the variables 
appearing in the Jost-Lehmann representation for $F_C(q)$ and $F_R(q)$ are Lorentz invariant objects. Thus going to a specific frame will not alter
the general attributes of the generalized functions. If we solve for $q_1^2$ in (\ref{nn20}) after obtaining an expression for $q_1^2$
\be
\label{nn21}
q_1=u_1\pm i\sqrt{\chi_{min}^2(u)-(q_0-u_0)^2+(q_2-u_2)^2+(q_3-u_3)^2+\rho}, \rho>0
\ee
We remind that the set of points$\{u_0,u_1,u_2,u_3 ; \chi_{min}^2= min~\chi^2 \}$ lie in $\bf S$. The above exercise has enabled us
to identify the domain where the singularities might lie with the choice for the variables $Q$ and $u$ we have made. We are dealing with
the equal mass case and note that the location of the singularities are symmetric with respect to the real axis. We now examine a further
simplified scenario where the coincidence region is bounded by two branches of hyperboloids so that ${\cal M}_+^2={\cal M}_-^2={\cal M}^2$.
Now the singular points are
\be
\label{nn22}
q_1=u_1\pm i\sqrt{Min~[\chi_{min}^2-u_0^2+u_2^2+u_3^2]+\rho},\rho>0
\ee
For the case under considerations:  $(Q+q)^2=(Q-q)^2={\cal M}^2$,
and  
\be
\label{nn23}
 q_1=u_1\pm i\sqrt{({\cal M}-\sqrt{Q^2-u_1^2)^2}+\rho},\rho>0
 \ee 
Now we can utilize this analysis to present a derivation of the Lehmann ellipse. The essential difference between the present investigation
in this context with the known results is that now we have to deal with several thresholds for identification of the coincidence
regions. These thresholds are the multiparticle states in various channels as discussed earlier as introduced in Section 2 through
the two equations (\ref{kk8}) and (\ref{kk9}). Their relevance is already reflected in the spectral representations, (\ref{nn10}) and (\ref{nn11}),
when we introduced complete set of intermediate states. We remark in passing that the presence of the excited KK states do not
shrink the singularity free regions. Therefore, the domain we have obtained is the smallest domain of analyticity; nevertheless, we feel
that in order to arrive at this conclusion the entire issue had to be examined with care.

\bigskip

\noindent {\it The Lehmann Ellipses}

\bigskip

\noindent  Our goal is to derive fixed-$t$ dispersion relations. We have noted that as $s\rightarrow s_{thr}$,   cos$\theta$ goes out of the
physical region $-1\le cos\theta \le +1$,  $\theta$  being  the $c.m.$ angle   when  we wish to hold $t$ fixed. We choose the following
kinematical configuration in order to derive the Lehmann ellipse. For the case at hand i.e. elastic scattering of equal (nonzero) charge KK
states, hence particles of equal mass. Here  $(a,b)$ and $(c,d)$ are respectively the incoming and outgoing particles. They are assigned the
following energies and momenta in the $c.m.$ frame: 
\be
\label{nn24}
p_a=(E_a,~{\bf k}),~~p_b=(E_b,-{\bf k}),~~p_c =(E_c,~{\bf k}'),~~p_d=(E_d,-{\bf k}')
\ee    
$\bf k$ is the $c.m.$ momentum, $|{\bf k}|=|{\bf k}'|$, $E_a=\sqrt{(m_a^2+{\bf k}^2)}$, $E_b=\sqrt{(m_c^2+{\bf k}^2)}$, $E_c=\sqrt{(m_c^2+{\bf k}'^2)}$ and
$E_d=\sqrt{(m_d^2+{\bf k}'^2)}$. Although all the particles, $(a,b,c,d)$,  are identical,  we keep labelling them as individual one for the purpose which
will be clear shortly. Thus $E_a=E_b$ and $E_c=E_d$ and ${\hat{\bf k}}.{\hat{\bf k}}'=cos\theta$.  It is convenient to choose the
following coordinate frame for the ensuing discussions.
\be
\label{nn25}
p_a=({\sqrt s},+{\bf k},~{\bf 0}),~~p_b=({\sqrt{s}},-{\bf k},~{\bf 0})
\ee
$\bf 0$ is the two spatial components of vector $\bf k$ and
\be
\label{nn26}
p_c=({\sqrt{s}}, +kcos\theta,+ksin\theta,0)~~p_d=({\sqrt s},-kcos\theta, -ksin\theta,0)
\ee
with $k=|{\bf k}|=|{\bf k}'|$.  Thus, $s=(p_a+p_b)^2=(p_c+p_d)^2$
\be
\label{nn27}
q={1\over 2}(p_d-p_c)=(0,-kcos\theta,-ksin\theta,0),~~P={1\over 2}(p_a+p_b)=({\sqrt s},0,0,0)
\ee
With these definitions of $q$ and $P$, when we examine the conditions for nonvanishing of the spectral representations of  $A_s$ and $A_u$ we arrive 
at 
\be
\label{nn28}
(P+q)^2>{{\cal M}_+}^2,~{\rm for}~A_s\ne 0,~~(P-q)^2>{{\cal M_-}^2},~{\rm for}~A_u\ne 0
\ee
Thus the coincidence region is given by the condition
\be
\label{nn29}
(P+q)^2<{{\cal M}_+}^2,~~(P-q)^2<{{\cal M}_-}^2
\ee
We are dealing with the equal mass case; therefore, ${{\cal M_+}}^2={{\cal M}_-}^2={\cal M}^2$. We conclude from the energy momentum 
conservation constraints  (use the expressions for $P$ and $q$) that $p_c^2=(P-q)^2<M_c^2$ and $p_d^2=(P+q)^2<M_d^2$ in this region.
Moreover, $(p_a-p_c)^2=(P-q-p_a)^2<{{\cal M}_{ac}}^2$ and $(p_a+p_d)^2=(P-q-p_a)^2<{{\cal M}_{ad}}^2$. We also note that $(P-q)\in V^+$ and
$(P+q)\in V^+$. The admissible hyperboloid is 
$(q-u)^2=\chi_{\min}^2+\rho, \rho>0$ with 
$({{p_a+p_b}\over 2}\pm u)\in V^+$.  $\chi_{min}^2$ assumes the following form for the equal mass case,
\be
\label{nn30}
\chi_{min}^2=Max~ \bigg\{0, {\cal M}-\sqrt{({{(p_a+p_b)}\over 2}+u)^2}, {\cal M}-\sqrt{{({(p_a+p_b)}\over 2}-u)^2} \bigg\}
\ee
Notice that ${\cal M}$ appearing in the second term of the curly in (\ref{nn30}) is the mass of two or more multiparticle states carrying the quantum
 numbers of particle $c$; whereas ${\cal M}$ appearing in the third term inside the curly bracket is the mass of two or more multiparticle states
 carrying the quantum numbers of particle $d$.
In the present case $\cal M$ has the same quantum number as that of the incoming state carrying KK charge $n$. Thus, in this sector, we can proceed to show 
the existence of the small Lehmann Ellips (SLE). It is not necessary to present the entire derivation here.  The extremum of the ellipse
is given by
\be
\label{nn31}
cos\theta_0=\bigg(1+{{(M_c^2-m_c^2)(M_d^2-m_d^2)}\over{k^2(s-M_c^2-M_d^2)}} \bigg)^{1/2}
\ee
We note that $M_c=\sqrt{m_n^2+m_0^2}$ is the mass of the lowest multiparticle state (one particle with KK charge $one$ and another with KK charge $zero$;
 moreover, $M_c=M_d$. Thus the denominator is $k^2s$. 
 \be
 \label{nn32}
cos\theta_0=\bigg(1+{{9m_0^4}\over{k^2s}}\bigg)^{1/2}
\ee
It will be a straightforward work to derive the properties of the large Lehmann Ellipse (LLE) by reducing all the four fields in the expression
for the four point function as is the standard prescription. also note that
 the value of $cos\theta (s)$ depends on $s$.
A natural question to ask is: what is the role of the KK towers?\\

\noindent {\it Important Remark:} The first point to note is that in the presence of the other states of KK tower, we have to carry out the same analysis as above for
each sector. Notice, however, each multiparticle state composed of KK towers has to have the quantum numbers of $c$ (same as $d$ since we
consider elastic channels of equal mass scattering). Thus if $c$ carries charge $n$, then a possible KK state could be $q+l+m=n$ since KK charges
can be positive and negative. The second point is when we derive the value of $cos\theta_0$, for each such case, it is rather easy to work out that
value will be away from original expression (\ref{nn31}). Thus the nearest singularity in $cos\theta$ plane is given by the expression
(\ref{nn32}) although there will be Lehmann ellipses associated with higher KK towers. 
\\
Consequently, when we expand the scattering amplitude in partial waves (in the Legendre polynomial basis) the domain of convergence is enlarged. This domain 
of analyticity is enlarged (earlier it was only physically permitted values of $cos\theta$) to a region which is an ellipse whose semimajor 
axis is given by (\ref{nn32}). Moreover, the absorptive part of the scattering amplitude has a domain of convergence beyond $cos\theta=\pm 1$;  it converges
inside the large Lehmann ellipse (LLE).
Therefore, we are able to write fixed-$t$ dispersion relations as long as $t$ lies in the following domain
\be
\label{nn33}
|t|+|t+4k^2|<4k^2cos\theta_0
\ee
The absorptive parts
$A_s$ and $A_u$ defined on the right hand and left hand cuts respectively, for
$s'>s_{thr}$ and $u'>u_{thr}$ are holomorphic in the LLE. Thus, assuming no
subtractions 
\be
\label{disperse}
F(s,t)={1\over \pi}\int _{s_{thr}}^{\infty}{{ds'~A_s(s',t)}\over{s'-s}}
+{1\over \pi}\int _{u_{thr}}^{\infty}{{du'~A_u(u',t)}\over{u'+s}-4m^2+t}
\ee
We shall discuss the issue of subtractions in sequel. We remark in passing that crossing has not been proved explicitly in this investigation.
However, it is quite obvious from the preceding developments, it will not be hard to prove crossing either from the prescriptions of
Bremmermann, Oehme and Taylor \cite{bot} or from the procedures of Bross, Epstein and Glaser \cite{beg1}.

\bigskip

\noindent  { \bf 4. Unitarity and Asymptotic Behavior of the Amplitude} 

\bigskip

\noindent In this section we shall explore the consequences of unitarity as mentioned earlier. The investigation so far has followed
what is known as {\it the linear program} in axiomatic field theory. All our conclusions about the analyticity properties of the scattering
amplitude are derived from micro causality, Lorentz invariance, translational invariance and axioms of LSZ. Note that unitarity of the
$S$-matrix is a nonlinear relationship and it is quite powerful. For example, the positivity properties of the partial wave amplitude
follows as a consequence. First we utilize unitarity in a new context in view of the fact that there are infinite towers of KK states
in the spectral representation of $F_C(q)$ and the representation for $F_R(q)$.  \\
  
  Let us define the $\bf T$-matrix as follows:
  \be
  \label{nn56}
  {\bf S}={\bf 1}-i{\bf T}
  \ee 
  The unitarity of the S-matrix, ${\bf {SS^{\dagger}}}={\bf {S^{\dagger}S}}={\bf 1}$ yields
  \be
  \label{nn57}
  ({\bf{T^{\dagger}}}-{\bf{T}})=i{\bf{T^{\dagger}T}}
  \ee
  In the present context,  we consider the matrix element  for the reaction $a+b\rightarrow c+d$. Note that on
  $L.H.S$ of ({\ref{nn5}) it is taken between ${\bf{T^{\dagger}-T}}$. We introduce a complete set of physical states
  between ${\bf {T^{\dagger}T}}$.   For the elastic case with all particles of KK charge, $n$, the unitarity relation is
  \bea
  \label{nn58}
  <p_d,p_c~in|{\bf{T^{\dagger}}}-{\bf T}|p_b,p_a~in>=i\sum_{{\cal N}}<p_d,p_c~in|{\bf{T^{\dagger}}}|{\cal N}><{\cal N}|{\bf T}|p_b,p_c~in>
  \eea
  The complete of states  stands for $|{\cal N}> =|{\cal P}_n{\tilde\alpha}_n>$.  The unitarity relation reads,
  \bea
  \label{nn59}
  T^*(p_a,p_b;p_c,p_d)-T(p_d,p_c;p_b,p_a)=2\pi i\sum_{\cal N}\delta(p_d+p_c-p_n)T^*(n;p_c,p_d)T(n;p_b,p_a)
  \eea
  We arrive at an expression like the second term of the $R.H.S$ of (\ref{nn1}) after reducing two fields. If we reduce a single field
  as the first step (as is worked out in text books) there will be a single KG operator acting on the field and eventually
  we obtain matrix element of only a single current.  The $R.H.S.$ of (\ref{nn7}) has matrix element like (for example)
  $p_a+p_b\rightarrow p_n$. Thus we can express it as \footnote{ We  adopt the arguments and procedures of Gasiorowicz in these derivations}
    \cite{gasio}
  \be
  \label{nn60}
  \delta(p_n-p_a-p_b)T(n:p_b,p_a)= (2\pi)^{3/2}<n~out|J_a(0)|p_b>\delta(p_n-p_a-p_b)
  \ee
  After carrying out the computations we arrive at
  \bea
  \label{nn61}
  T(p_d,p_c;p_b,p_a) -T^*(p_d,p_c;p_b,p_a)=&&\sum_{\cal N}\bigg[\delta(p_d+p_c-p_n)\times \nonumber\\&& T(p_d,p_c;n)T^*(n;p_b,p_a)-\nonumber\\&&
  \delta(p_a-p_c-p_n)\times \nonumber\\&& T(p_d,-p_c;n)T^*(p_d,-p_c;n) \bigg]
  \eea
  Let consider the scattering amplitude for the reaction under considerations.
  \bea
  \label{nn62}
  F(s,t)=i\int d^4xe^{i(p_a+p_c).{{x}\over 2}}\theta(x_0)<p_d|[J_a(x/2),J_c(-x'/2)]|p_b>
  \eea
  We evaluate the imaginary part of this amplitude, $F(s,t)$
  \bea
  \label{nn63}
  Im~F(s,t)=&&{{1}\over{2i}}(F-F^*)\nonumber\\&&
  ={{1}\over2}\int d^4xe^{i(p_a+p_c).{{x\over 2}}}<p_d|[J_a(x/2),J_c(-x/2)]|p_b>
  \eea
  Note that $F^*$ is invariant under interchange $p_b\rightarrow p_d$ and also $p_d\rightarrow p_b$; moreover, $\theta(x_0)+\theta(-x_0)=1$.
  We open up the commutator of the two currents in (\ref{nn11}). Then introduce a complete set of physical states $\sum_{\cal N}|{\cal N}><{\cal N}|=1$. Next we
  implement  translation operations in each of the (expanded) matrix elements to  express  arguments of each current as  $J_a(0)$ and $J_c(0)$
  and finally integrate over $d^4x$  to get the $\delta$-functions. As a consequence  (\ref{nn62}) assumes the form
  \bea
  \label{nn65}
  F(p_d,p_c;p_b,p_a)-F^*(p_b,p_a;p_c,p_d)=&&2\pi i\sum_{\cal N}\bigg[\delta(p_d+p_c-p_n)F(p_d,p_c;n)F^*(p_a,p_b;n)\nonumber\\&&
  -\delta(p_a-p_c-p_n)\times \nonumber\\&& F(p_d,-p_a;n)F^*(p_b,-p_c;n)\bigg]
  \eea
  This is the generalized unitarity relation where all external particles are on the mass shell. Notice that the first term on the $R.H.S$ of the
  above equation is identical in form to the $R.H.S.$ of (\ref{nn9}); the unitarity relation for ${\bf T}$-matrix. The first term in (\ref{nn65})  has
  the following interpretation: the presence of the $\delta$-function and total energy momentum conservation implies
  $p_d+p_c=p_n=p_a+p_b$. We identify it as the $s$-channel process $p_a+p_b\rightarrow p_c+p_d$.\\
  Let us examine the second term of (\ref{nn65}).  Recall that the unitarity holds for the $S$-matrix when all external particles are on shell
  (as is true for the $T$-matrix). The presence of the $\delta$-function in the expression ensures that the intermediate physical states will contribute for
  \be
  \label{nn66}
  p_b+(-p_c)=p_n=p_d+(-p_a)
  \ee
  The masses of the intermediate states must satisfy
  \be
  \label{nn67}
  {\cal M}_n^2=p_n^2=(p_b-p_c)^2
  \ee  
  It becomes physically transparent if we choose the Lorentz frame where  particle  $'b'$ is at rest i.e. $p_b=(m_b, {\bf 0})$; thus
  \be
  \label{nn68}
  {\cal M}_n^2=2m_b(m_b-p_c^0),~~p_c^0>0
  \ee
  since $m_b=m_c$ and $p_c^0={{\sqrt{m_c^2+{\bf p}_c^2}}}={{\sqrt{m_b^2+{\bf p}_c^2}}} $;  ${\cal M}_n^2<0$ in this case.
  We recall that all particles carry KK charge $n$ and hence the mass is $m_b^2=m_n^2=m_0^2+{{n^2}\over{R^2}}$. The intermediate state
  must carry that quantum number. In conclusion, the second term of (\ref{nn65}) does not contribute to the $s$-channel reaction. There is
  an important implication of the generalized unitarity equation: Let us look at the crossed channel reaction
  \be
  \label{nn69}
  p_b+(-p_c)\rightarrow p_d+(-p_a);~~ -p_a^0>0,~ and~ -p_c^0>0
  \ee
  Here $p_b$ and $p_c$ are incoming (hence the negative sign for $p_c$) and $p_d$ and $p_a$ are outgoing. The second matrix element
  in (\ref{nn65}) contributes to the above process in the configurations of the four momenta of these particle;
   whereas the first term in that equation does not if we follow the arguments for
  the $s$-channel process. \\
  {\it Remark}: We notice the glimpses of crossing symmetry here.  Indeed, the starting
  point will be to define $F_C(q)$ and look for the coincidence region. Notice that  $q$ is related to physical momenta of external particles when
  $|Q_i>$ and $|Q_f>$ are identified with  the momenta of the 'unreduced' fields. Indeed, we could proceed to prove crossing  symmetry for the 
  scattering process; however,  it is not our present goal.\\
   {\it An important observation is  in order}: \\
     We could ask whether entire Kaluza-Klein tower  of states
  would appear as intermediate states in the unitarity equation.
  It is obvious from the unitarity equation (\ref{nn65}) that for the $s$-channel process, due to the presence of the energy momentum
  conserving $\delta$-function, $p_n^2={\cal M}_n^2= (p_a+p_b)^2$; consequently, not all states of the infinite KK tower will contribute
  to the reaction in this,   ($s$), channel. Therefore the sum would terminate after finite number of terms, even for very large $s$ as long as it is finite.
   Same argument also holds for the crossed channel process. Thus unitarity constraint settles the issue of the contributions of KK towers
   as we alluded to in the previous section in the context of the spectral representation of $F_R(q)$, $F_A(q)$ and $F_C(q)$.  
  \\
  Let us turn the attention to the partial wave expansion of the amplitude and the power of the positivity property of absorptive part of the
  amplitude. We recall that the scattering amplitude admits a partial wave expansion
  \be
  \label{uni1}
  F(s,t)= {{\sqrt s}\over k} \sum_{l=0}^{\infty}(2l+1)f_l(s)P_l(cos\theta)
  \ee
  where $k=|{\bf k}|$ and $\theta$ is the $c.m.$ scattering angle. The expansion converges inside the Lehmann ellipse with with
  focii at $\pm 1$ and semimajor axis $1+{{const}\over {2k^2}}$. Unitarity  leads to the positivity constraints on the partial wave
  amplitudes
  \be
  \label{uni2}
  0\le |f_l(s)|^2\le Im~f_l(s)\le 1
  \ee
  As is well known, the semimajor axis of the Lehmann ellipse shrinks as $s$ grows. Recall that derivation of the Lehmann ellipse is
  based on the {\it linear program}.  Martin \cite{martin1}  has proved an important theorem. It is known as the procedure for the enlargement of the domain of analyticity.
  He demonstrated that the scattering amplitude is analytic in the topological product of the domains $D_s\otimes D_t$. This domain is
  defined by $|t|<{\tilde R}$, ${\tilde R}$ being independent of $s$ and $s$ is outside the cut $s_{thr}+\lambda=4m_n^2+\lambda, \lambda>0$. In order
  to recognize the importance of this result, we briefly recall the theorem of BEG \cite{beg2}. It is essentially  the study of the analyticity property of the
  scattering amplitude $F(s,t)$. It was shown that in the neighborhood of any point $s_0$, $t_0$  $ -T<t_0\le 0$, $s_0$ outside the cuts, there is
  analyticity in $s$, and $t$ in a region
  \be
  \label{uni3}
  |s-s_0|<\eta_0(s_0,t_0),~~|t-t_0|<\eta_0(s_0,t_0)
  \ee
  The amplitude is analytic.
  Note the following features of BEG theorem: it identifies the domain of analyticity; however, the size of this domain may vary as $s_0$ and $t_0$
  vary. Furthermore, the size of this domain might shrink to zero; in other words, as $s\rightarrow 0$, $\eta(s)$ may tend to zero. The importance Martin's
  theorem lies in his proof that $\eta(s)$  is bounded from below i.e. $\eta (s)\ge {\tilde R}$ and ${\tilde R}$ is $s$-independent. It is unnecessary to repeat
  the proof of Martin's theorem here. Instead, we shall summarize the conditions to be satisfied by the amplitude as stated by Martin \cite{martin1}.
  \\
  {\it Statement of Martin's Theorem:} 
   {If following
requirements are satisfied by the elastic amplitude\\
I. $F(s,t)$ satisfies fixed-t dispersion relation in s with finite number of
subtractions ($-T_0\le t\le 0$).\\
II. $ F(s,t)$ is an analytic function of the two Mandelstam variables, $s$ and
$t$, in a neighborhood of $\bar s$ in an interval below the threshold,
$4m_n^2-\rho<{\bar s}<4m_n^2$ and also in some neighborhood of $t=0$,
$|t|<R({\bar s})$. This statement hold due to the work of
Bros, Epstein and Glaser \cite{beg1,beg2}.\\
III. Holomorphicity of $A_s(s',t)$ and $A_u(u',t)$: The absorptive parts of
$F(s,t)$ on the right hand and left hand cuts with $s'>4m_n^2$ and $u'>4m_n^2$
are holomorphic in the LLE. \\
IV. The absorptive parts $A_s(s',t)$ and $A_u(u',t)$, for $s'>4m_n^2$ and
 $u'>4m_n^2$ satisfy the following positivity properties
\bea
\label{uni4}
\bigg|{\bigg({\partial\over{\partial t}}}A_s(s',t)\bigg)^n\bigg|
\le {\bigg({\partial\over{\partial t}}\bigg)^n} A_s(s',t)\bigg|_{t=0},~~
-4k^2\le t\le 0
\eea
and
\bea
\label{uni5}
\bigg|{\bigg({\partial\over{\partial t}}}A_u(u',t)\bigg)^n\bigg|
\le {\bigg({\partial\over{\partial t}}\bigg)^n} A_u(u',t)\bigg|_{t=0},~~
-4k^2\le t\le 0
\eea
where ${\bf k}$ is the {\it c.m. }momentum.
Then $F(s,t)$ is analytic in the quasi topological product of the domains $D_s\otimes D_t$. (i) $s\in~cut-plane$: $s\ne 4m_n^2+\rho,\rho>0$ and
(ii) $|t|<{\tilde R}$, there exists some ${\tilde R}$ such that dispersion relations are valid for $|t|<{\tilde R}$, independent of $s$.
We may follow the standard method to determine ${\tilde R}$. The polynomial boundedness, in $s$,  can be asserted by invoking the simple arguments
 presented earlier. Consequently, a dispersion relation can be written down for $F(s,t)$ in the domain $D_s\otimes D_t$.  The importance of Martin's
 theorem is appreciated from the fact that it implies that the $\eta$ of BEG is bounded from below by an $s$-independent $\tilde R$. Moreover, value
 of $\tilde R$ can be determined by the procedure of Martin (see \cite{sommer} for the derivations). \\
 We shall list a few more results as corollary without providing detailed computations: \\
 (i) It can be proved that the partial wave expansion can be expressed as sum of two terms
 \be
 \label{uni6}
 F(s,t)= {\cal S}_1+{\cal S}_2
 \ee
 \be
 \label{uni6}
 {\cal S}_1={{\sqrt s}\over k}\sum_{l=0}^L(2l+1)f_l(s)P_l(1+{{t\over{2k^2}}})
 \ee
 \be
 \label{uni7}
 {\cal S}_2={{\sqrt s}\over k}\sum_{L+1}^{\infty}(2l+1)f_(s)P_l(1+{{t\over{2k^2}}})
 \ee
 where $L=const. {\sqrt s}log s$ is the cut off which is derived from the convergence of the partial wave expansion inside
 the Lehmann-Martin ellipse and the polynomial boundedness of the amplitude. The partial sum ${\cal S}_2$ has subleading 
 contributions to the amplitude compared to ${\cal S}_1$; in fact ${{\cal S}_1\over{{\cal S}_2}}\rightarrow (log~s)^{-1/4}$ for asymptotic
 $s$ apart from some innocent $t$-dependent prefactor; as is well known. \\
 (ii) {\it The Bound on $\sigma_t$}: The analog of Froissart-Martin bound can be obtained in that
 \bea
  \sigma_t(s)\le const. (log~s)^2
  \eea
  The constants appearing in the determination of 
 $L$ and in derivation of the Froissart-Martin bound can be determined in terms of $\tilde R$ and we have refrained from giving those
 details here.\\
 (iii){\it Number of subtractions}:  Once we have derived (i) and (ii) it is easy to prove the Jin-Martin \cite{jml}  bound which states that the amplitude requires at most two subtractions.
 This is achieved by appealing to the existence of fixed-$t$ dispersion relation relations and to Phragman-Lindelof theorem.  \\
 We would like to draw the attention of the reader to the fact  that  a field theory defined on the manifold $R^{3,1}\otimes S^1$ whose spectrum consists
 of a massive scalar field and a tower of Kaluza-Klein states satisfies nonforward dispersion relations. This statements begs certain clarifications.
 The theory satisfies LSZ axioms. The analyticity properties can be derived in the {\it linear program} of axiomatic field theory which
 leads to the proof of the existence of the Lehmann ellipses. The role of the KK tower is to be assessed in this program. Once we invoke
 unitarity constraint stronger results follow and the enlargement of the domain of analyticity in $s$ and $t$ variables can be established.

\bigskip

\noindent{\bf 5 Summary and Discussions}

\bigskip

\noindent We summarize our results in this section and discuss their implications.  The objective of the present work is to investigate
the analyticity property of the scattering amplitude in a field theory with a compactified spatial dimension on a circle i.e. the $S^1$
compactification. We were motivated to undertake this work from work of Khuri \cite{khuri2} who considered
potential scattering with a compact spatial coordinate. He showed the lack of analyticity of the forward scattering amplitude under certain 
circumstances. Naturally, it is important to examine what is the situation in relativistic field theories. As has been emphasized by us before,
lack of analyticity of scattering amplitude in a QFT will be a matter of concern since analyticity is derived under very general
axioms of  QFT. Thus a compactified spatial coordinate in a theory with flat Minkowski spacetime coordinates should not lead
to unexpected drastic violations of fundamental principles of QFT. In this paper, initially, a five dimensional neutral massive scalar theory of mass, $m_0$, 
was considered in a flat 
Minkowski spacetime.  Subsequently, we compactified a spatial coordinate on $S^1$ leading to a spacetime manifold $R^{3,1}\otimes S^1$.
The particles of the resulting theory are a scalar of mass $m_0$ and the Kaluza-Klein towers. In this work,
we have focused on elastic scattering of states carrying nonzero equal KK charges, $n\ne 0$ to prove fixed-$t$ 
dispersion relations. We have left out the elastic scattering of $n=0$ states as well as elastic scattering of an $n=0$ state
with an $n\ne 0$ state for nonforward directions. These two cases can be dealt with without much problem from our present
work. Moreover, our principal task is to prove analyticity for scattering of $n\ne 0$ states and thus complete
the project we started with in order to settle the issue related to analyticity as was raised by Khuri  \cite{khuri2} in the context of potential scattering. 
We showed in I that forward amplitude satisfies
dispersion relations. However, it is not enough to prove only the dispersion relations for the forward  amplitude but a fixed-$t$
dispersion relation is  desirable. We have adopted the LSZ axiomatic formulation, as was the case in I, for this purpose. Our results, consequently,
do not rely on perturbation theory whereas, Khuri \cite{khuri2} arrived at his conclusions in the perturbative Greens function techniques
as  suitable for a nonrelativistic potential model. Thus the work presented here, in some sense, has explored more than 
what Khuri had investigated  in the potential scattering. \\
We have gone through several steps, as mentioned in the discussion section of I, in order to accomplish our goal. 
The principal results of this are as follows. First we obtain a spectral
representation for the Fourier transform of the causal commutator, $F_c(q)$. We discussed the coincidence region which is important
for what followed. In order to identify the singularity free domain, we derived analog of the Jost-Lehmann-Dyson theorem. A departure from
the known theorem is that there are several massive states, appearing in the spectral representation, and their presence has to be
taken into considerations. Thus, we identified the the singularity free region i.e. the boundary of the domain of analyticity. Next, we derived
the existence of the Lehmann ellipses. We  were able to write down fixed-$t$ dispersion relations for $|t|$ within the Lehmann ellipse.\\
We have proceeded further. It is not enough to obtain the Lehmann ellipse since the semimajor axis of the ellipse shrinks as $s$ increases.
Thus it is desirable to derive the analog of Martin's theorem \cite{martin1}. We appealed to unitarity constraints following Martin and utilized his
arguments on the attributes of the absorptive amplitude and showed that indeed Martin's theorem can be proved for the case at hand.
As a consequence, the analog of Froissart-Martin upper bound on total cross sections, for the present case, is obtained. The convergence
of partial wave expansions within the Lehmann-Martin ellipse and polynomial boundedness for the amplitude, $F(s,t)$ for $|t|$ within Lehmann-Martin
ellipse, lead to the Jin-Martin upper bound \cite{jml} for the problem we have addressed here. In other words, the amplitude, $F(s,t)$,  does not need more than two 
subtractions   to write fixed $t$ dispersion relations for in the domain $D_s\otimes D_t$.  \\
We have made two assumptions: (i)  existence of stable particles in the entire spectrum of the
the theory defined on $R^{3,1}\otimes S^1$  geometry. Our arguments is based on the conservation of KK discrete charge $q_n={{n}\over R}$; 
it is the momentum along the compatified direction.(ii)  The 
absence of bound states.  We have presented some detailed arguments in support of (ii).  To put is very concisely, we conveyed that this
flat space  $D=4$
theory with an extra compact $S^1$ geometry results from toroidal compactification of five dimensional defined in flat Minkowski space.
In absence of gravity in $D=5$, the lower dimensional theory would not have massless gauge field and consequently, BPS type 
states are absent. It is unlikely that the massive scalars (even with KK charge) would provide bound states. This is our judicious conjecture. \\
Another interesting aspect needs further careful consideration. Let us start with a five dimensional Einstein theory minimally coupled to a massive
neutral scalar field of mass $m_0$. We are unable to fulfill requirements of LSZ axioms in the case of the five dimensional theory in curved spacetime. Furthermore,
let us compactify this theory to a geometry $R^{3,1}\otimes S^1$. Thus the resulting scalar field lives in flat Minkowski space with a compact
dimension. We have an Abelian gauge field in $D=4$, which arises from $S^1$ compactification of the 5-dimensional Einstein metric. The spectrum
of the theory can be identified: (i) There is a massive scalar of mass $m_0$ descending  from  $D=5$ theory accompanied by KK tower of states.
(ii) A massless gauge boson and its massive KK partners. (iii) If we expand the five dimensional metric around four dimensional Minkowski
metric when we compactify on $S^1$, we are likely to have massive spin 2 states (analog of KK towers). We may construct a Hilbert space
in $D=4$ i.e with geometry $R^{3,1}\otimes S^1$.  It will be interesting to investigate the analyticity properties of the scattering amplitudes and
examine the high energy behaviors. Since only a massless spin 1 particle with Abelian gauge symmetry appears in the spectrum, it looks as
if the analyticity of amplitudes will not be affected. However, there might be surprises since a massive spin 2 particle is present in the spectrum. 
Khuri  \cite{khuri2} was motivated by the large extra dimension scenario to undertake the problem.  He had raised the question what will be the consequences
of his conclusions (in the potential scattering model) if indeed the dispersion relation is not valid at LHC energies. However, the field theory
we have considered here,  the  dispersion relations are proved for fixed $t$. It will be worthwhile to undertake phenomenological analysis to check if there
are Froissart-Martin bound violation at extremely high energies. If there is a deviation from the $(ln~s)^2$ fit to $\sigma_t(s)$ for the available
data for $\sigma_t$ in the Tevatron and LHC energies, then one might (possibly) attribute 
such a violation to decompactification of a spatial dimension rather than cast doubts on  the fundamental principles of relativistic quantum field theories.
    
\bigskip

\noindent { \bf Acknowledgements}: It is my great pleasure to thank Andr\'e Martin who inspired me to undertake this investigation. I am grateful to him
for numerous valuable discussions. I wish to  thanks  Stefan Theisen for his encouragements and for discussions.  The gracious hospitality of Hermann Nicolai and the Max-Planck Instit\"ut  f\"ur Gravitational
Physik and the Albert Einstein Instit\"ut is gratefully acknowledged.

\end{document}